\documentclass[symmetry,article,accept,moreauthors,pdtex,a4paper]{Definitions/mdpi} 

\firstpage{1} 
\pubvolume{11}
\issuenum{9}
\articlenumber{1090}
\pubyear{2019}
\copyrightyear{2019}
\history{Received: 7 August 2019; Accepted: 21 August 2019; Published: 1 September 2019}
\updates{yes} 

\usepackage{amsfonts}
\usepackage{amssymb}
\usepackage{upgreek}

\Title{Faraday and Resonant Waves in Dipolar Cigar-Shaped Bose-Einstein Condensates}

\Author{Du\v{s}an Vudragovi\'{c} \orcidA{} and Antun Bala\v{z} *\orcidB{}}

\AuthorNames{Dusan Vudragovic and Antun Balaz}

\address[1]{%
Scientific Computing Laboratory, Center for the Study of Complex Systems, Institute of Physics Belgrade, University of Belgrade, Pregrevica 118, 11080 Belgrade, Serbia}

\corres{\hangafter=1 \hangindent=1.0em \hspace{-1em} Correspondence: antun.balaz@ipb.ac.rs
}

\abstract{Faraday and resonant density waves emerge in Bose-Einstein condensates as a result of harmonic driving of the system. They represent nonlinear excitations and are generated due to the interaction-induced coupling of collective oscillation modes and the existence of parametric resonances. Using a mean-field variational and a full numerical approach, we studied density waves in dipolar condensates at zero temperature, where breaking of the symmetry due to anisotropy of the dipole-dipole interaction (DDI) plays an important role. We derived variational equations of motion for the dynamics of a driven dipolar system and identify the most unstable modes that correspond to the Faraday and resonant waves. Based on this, we derived the analytical expressions for spatial periods of both types of density waves as functions of the contact and the DDI strength. We compared the obtained variational results with the results of extensive numerical simulations that solve the dipolar Gross-Pitaevskii equation in 3D, and found a very good agreement.}

\keyword{Bose-Einstein condensate; pattern formation; dipole-dipole interaction; parametric resonance; interaction effects}

\begin{document}

\section{Introduction}
\label{sec:introduction}

After pioneering experiments that realized Bose-Einstein condensates (BEC) in systems with weak contact interactions, it took a decade of work on improvements of experimental techniques to enable measurement of effects of the dipole-dipole interaction (DDI) that exist between atoms or molecules with a permanent or induced electrical or magnetic dipole moment. The very first such experiment was realized in 2005 with chromium atoms $^{52}$Cr~\cite{Nature.448.672}, followed by the experiments with atoms with much larger magnetic moments, such as dysprosium $^{164}$Dy~\cite{PhysRevLett.107.190401} and erbium $^{168}$Er~\cite{PhysRevLett.108.210401}. Furthermore, the~dipolar BECs comprised of polar molecules with much stronger electrical~\cite{NatPhys.7.502} and magnetic~\cite{PhysRevLett.115.203201} dipole moments were also realized. While the contact interaction is symmetric and has a short-range, the DDI between atoms or molecules is anisotropic and long-range. These features are responsible for a whole series of new phenomena that appear in ultracold dipolar gases~\cite{PhysRep.464.71}. If we take into account that the strength of the contact interactions can be varied over many orders of magnitude using the Feshbach resonance~\cite{Nature.392.32354} technique, and that the DDI strength can be also tuned using a fast rotating magnetic or electric field~\cite{PhysRevLett.89.130401,PhysRevLett.120.230401}, it is easy to see that such a versatility of dipolar quantum gases is unparalleled and makes them an obligatory element in a toolbox for engineering quantum devices and sensors.

Bose-Einstein condensates are usually termed quantum fluids, which encompasses a broader range of physical systems where quantum effects are either dominant or very much pronounced. Despite their name, some of quantum fluids do not share the trademark property of classical fluids, incompressibility. In fact, the BECs are made of rarefied gases, but their fluid-like behavior stems from the quantum coherence of such systems. Therefore, while in classical fluids density modulations can be excited only under extreme conditions, in quantum fluids the density waves represent one of the natural collective excitations. They appear due to nonlinearity in ultracold quantum gases, and~can be induced by a harmonic modulation of the trap frequencies or interaction strengths. 
However, the~motivation for study of such excitations comes from the classical phenomenon of Faraday waves, which may appear on the surface of the shallow layer of liquid under certain conditions. Namely, if~the container with the liquid is harmonically oscillated in a vertical direction, the wave patterns may emerge, depending on the ratio of the liquid depth and the container size, as well as depending on the modulation frequency. This phenomenon was first studied and described by Michael Faraday at the beginning of  19th century~\cite{PhilTransRSocLond.121.299}. The interest for this type of excitations arose again during the 1980s, as a consequence of the study of nonlinear liquids. In the context of ultracold gases, Faraday waves were first investigated theoretically in 2002 by Staliunas~\cite{PhysRevLett.89.210406}. After these theoretical and numerical results for the systems with contact interaction, where it was assumed that the interaction strength is harmonically modulated, the Faraday waves were first measured in BEC experiments with $^{87}$Rb in 2007 by Engels~\cite{PhysRevLett.98.095301}, and more recently with $^{7}$Li by Hulet and Bagnato~\cite{PhysRevA.81.053627,PhysRevX.9.011052}. In the first experiment,  the radial part of the harmonic trap was modulated, while the other two experiments have modulated the contact interaction strength. However, qualitatively, this leads to the same type of density waves.

Parametric driving of system parameters can lead to pattern formation not only in BECs, where Faraday waves are experimentally observed in cigar-shaped condensates~\cite{PhysRevLett.98.095301,PhysRevA.81.053627,PhysRevX.9.011052}, but also in helium cells~\cite{PhysRevE.76.046305}. The actual experimental observation of this phenomenon in 2007 was preceded by numerical studies starting in 2002~\cite{PhysRevLett.89.210406,PhysRevA.76.063609,PhysicaA.389.4663,ProcRomAcad.12.209,RomRepPhys.63.1329,RomRepPhys.65.820}, all focusing on systems with short-range, contact interactions. More recently, Faraday waves have been studied in dipolar~\cite{PhysRevA.81.033626,PhysRevA.86.023620,ProcRomAcad.14.35} and two-component condensates, including the systems with spatially-dependent contact interaction~\cite{PhysRevA.85.023613,PhysRevA.89.023609}. Numerical studies of Faraday waves have also been extended to mixtures of Bose and Fermi gases~\cite{PhysRevA.87.023616}, as well as Fermi gases exhibiting superfluid behavior~\cite{PhysRevA.78.043613,JPhysB.44.115303}.

Faraday waves in ultracold gases are a consequence of the existence of parametric resonances in the system. While the spatial period of these waves depends on the geometry of the system and other parameters, the frequency of their oscillations is constant and is two times smaller than the modulation frequency. This is a characteristic of all parametric resonant phenomena, and in the variational approach leads to the Mathieu's differential equation~\cite{McLachlan}, which gives the observed ratio of the frequency of Faraday waves and the modulation. The Faraday density waves with half of the modulation frequency are not the only nonlinear excitation of the system. In a driven system,  there are always excitations with the same frequency as the modulation. However, they become resonant when the modulation frequency corresponds to one of the collective modes or the trap frequencies, or their linear combination. The resonant waves develop in the system and grow exponentially~\cite{PhysRevE.84.056202}, faster than the Faraday waves. Therefore, these two phenomena can be easily distinguished, not only by comparing their frequencies, but also the corresponding onset times. We note that resonant behavior can appear not only due to the modulation of the interaction strength or the trapping potential, but~also due to its spatial modulation~\cite{PhysRevA.83.013603,PhysRevA.84.013626,PhysRevA.84.053627,PhysRevA.87.015601,PhysRevA.85.033635,PhysRevA.85.062110,PhysRevA.88.033621,PhysRevA.85.013630,QuantumInfProcess.12.3569,JLowTempPhys.173.177}.

In the context of dipolar BECs, the study of Faraday waves was limited mostly to their excitation spectrum in one-dimensional and two-dimensional systems~\cite{PhysRevA.81.033626}, while the properties of resonant waves, to the best of our knowledge, have not been studied  yet. In Section~\ref{sec:variational}, we develop a mean-field variational approach for the dynamics of a driven dipolar BEC at zero temperature and identify the instability of the system leading to the emergence of Faraday and resonant waves. Using this approach, we derive analytic expressions for the dependence of density wave properties on the strength of the contact and the dipole-dipole interaction. In Section~\ref{sec:faraday}, we numerically study how such waves develop and can be characterized in ultracold systems of three experimentally relevant magnetic dipolar species: chromium $^{52}$Cr, erbium $^{168}$Er, and dysprosium $^{164}$Dy. In Section~\ref{sec:interaction}, the analytically obtained expressions for the spatial period of Faraday are compared to results of the extensive numerical simulations, which solve the full three-dimensional mean-field equations for a dipolar BEC. The emergence of resonant waves and comparison of the corresponding analytical and numerical results is given in Section~\ref{sec:resonant}. Finally, Section~\ref{sec:conclusions} summarizes our conclusions and presents outlook for future research.

\section{Variational Approach}
\label{sec:variational}

We consider the system in an experimentally-inspired setup, where the condensate is confined into a cigar-shaped harmonic trap, with the equilibrium frequencies $\omega_{x} = 2 \pi \times 7$~Hz, $\omega_{y} =\omega_{z} = \Omega_0 = 2 \pi \times 160.5$~Hz. These are typical values taken from Reference~\cite{PhysRevLett.98.095301}. The dipole moments of the atoms are assumed to be oriented along $z$ direction, i.e., orthogonal to the weak-confinement axis $x$ (which we refer to as the longitudinal axis), since this maximizes the stability of the system. To~ensure stability of the system, we consider the condensate to have $N=10^4$ atoms for all three species. The~driving of the system is achieved by harmonic modulation of the radial ($y-z$) part of the trap,
\begin{equation}
    \omega_y(t) = \omega_z(t) = \Omega_0\,  (1 + \epsilon \sin \omega_m t) \, , 
    \label{eq:faraday_omega_yz}
\end{equation}
where $\epsilon = 0.1 - 0.2$ is the modulation amplitude and $\omega_m$ is the modulation frequency.

For a variational study of Faraday and resonant waves in dipolar condensates, we use a modification of the Gaussian ansatz~\cite{RomRepPhys.65.820,JPhysB.49.165303,PhysicaA.391.1062,PhysRevA.76.063609,PhysicaA.389.4663,ProcRomAcad.12.209,RomRepPhys.63.1329,ProcRomAcad.14.35,PhysRevA.89.023609,PhysRevE.84.056202,PhysRevA.85.023613} to capture the induced density waves in the longitudinal, weak-confinement direction $x$,
\begin{equation}
    \psi(x, y, z, t) = A \, e^{
        - \frac{x^2}{2 u_x^2} - \frac{y^2}{2 u_y^2} - \frac{z^2}{2 u_z^2}
        + i x^2 \phi_x + i y^2 \phi_y + i z^2 \phi_z
    } \left[ 1 + (\alpha + i \beta) \cos kx \right] \, ,
    \label{eq:faraday_ansatz}
\end{equation}
where the normalization of the wave function to unity is ensured by the prefactor
\begin{equation}
   A \equiv A(u_x, u_y, u_z, \alpha, \beta, k) =
    \frac{1}{\pi^{3/4} \sqrt{u_x u_y u_z}}
    \frac{\sqrt{2}}
         {\sqrt{2 + \alpha^2 + \beta^2 + 4 \alpha \, e^{- k^2 u_x^2 / 4} + 
                (\alpha^2 + \beta^2) \, e^{- k^2 u_x^2}}} \, .
    \label{eq:faraday_ansatz_norm}
\end{equation}

The above variational ansatz involves eight variational parameters $\{ u_i, \phi_i, \alpha, \beta\}$, which are functions of time. The parameters $u_i$ represent the condensate widths, while $\phi_i$ are the conjugated phases, which are necessary to properly describe the system's dynamics. Note that these phases can be omitted when we are interested only in the ground state. The multiplicative factor $1 + (\alpha + i \beta) \cos kx$ describes the density modulation along $x$ direction, and the variational parameters $\alpha$ and $\beta$ represent the real and the imaginary part of the amplitude of the wave. The wave vector $k$, which is related to the spatial period $\ell$ of the density waves by $\ell = 2 \pi / k$, is not treated here as a variational parameter. We determine its value from the condition for the instability emergence, which leads to Faraday or resonant waves.

The use of the Gaussian variational ansatz corresponds to the weak interaction regime with low density of atoms, while the Thomas-Fermi profile is more appropriate for systems with high particle density. Although the emergence of Faraday and resonant waves leads to higher particle densities, we still use the Gaussian ansatz in all regimes. This is done since we are mostly interested just in the onset of longitudinal density modulations, but also for mathematical convenience. Let us note that tunability of all variational parameters may improve the accuracy of the applied approximation. Nevertheless, the use of this ansatz can be fully justified only {      a posteriori}, by comparison with numerical results  \cite{PhysRevA.76.063609}.

Note that we use the dimensionless units, where a chosen referent frequency $\omega_r$ defines the length scale through the harmonic oscillator length $\sqrt{\hbar / (m \omega_r)}$, where $m$ is the mass of the corresponding atomic species, the time scale as $1 / \omega_r$, and the energy scale as $\hbar \omega_r$. The trapping frequencies are also expressed in units of $\omega_r$ through the trap aspect ratios $\gamma = \omega_x / \omega_r$, $\nu = \omega_y / \omega_r$, and $\lambda = \omega_z / \omega_r$, as well as the modulation frequency $\eta_m=\omega_m/\omega_r$. We   choose below the value $\omega_r=\Omega_0$, corresponding to $\nu=\lambda=1$, but for now we keep all three aspect ratios as free parameters, for generality.

If we insert the modified Gaussian ansatz (Equation \eqref{eq:faraday_ansatz}) into the Lagrangian density that yields the dipolar Gross-Pitaevskii equation, we can express the Lagrangian of the system as a sum of five terms. The~first term reads
\begin{equation}
    L_1(t) = \frac{i}{2} \int d{\bf r} \, \left(\psi^* \dot \psi - \psi \dot \psi^* \right)
           = - \frac{1}{2} \left(u_x^2 \dot \phi_x + u_y^2 \dot \phi_y + u_z^2 \dot \phi_z \right) 
             - \frac{\alpha \dot \beta - \beta \dot \alpha}{2 + \alpha^2 + \beta^2} \, ,
    \label{eq:faraday_lagrangian_l1}
\end{equation}
while the kinetic and the potential energy terms yield, respectively,
\begin{gather}
    L_2(t) = \frac{1}{2} \int d{\bf r} \, \psi^* \Delta \psi
           = - \frac{1}{4} \left(
                \frac{1}{u_x^2} + \frac{1}{u_y^2} + \frac{1}{u_z^2} + 
                4 u_x^2 \phi_x^2 + 4 u_y^2 \phi_y^2 + 4 u_z^2 \phi_z^2
                \right)
              - \frac {(\alpha^2 + \beta^2) \, k^2}{2 (2 + \alpha^2 + \beta^2)} \, ,
    \label{eq:faraday_lagrangian_l2}\\
    L_3(t) = - \int d{\bf r} \, \left(\frac{1}{2}\gamma^2 x^2+\frac{1}{2}\nu^2 y^2 + \frac{1}{2}\lambda^2 z^2\right) \left| \psi \right|^{2} = - \frac{1}{4} 
            \left( \gamma^2 u_x^2 + \nu^2 u_y^2 + \lambda^2 u_z^2 \right) \, .
    \label{eq:faraday_lagrangian_l3}
\end{gather}

The contact interaction term corresponds to
\begin{equation}
    L_4(t) = - 2 \pi N a_s \int d{\bf r} \, \left| \psi \right|^{4} =
             - \frac{N a_s}{\sqrt{2 \pi} \, u_x u_y u_z} \left(1 + 
             \frac {\alpha^4 + 16 \alpha^2 + 2  \alpha^2 \beta^2 + \beta^4)}{2 (2 + \alpha^2 + \beta^2)^2} \right) \, ,
    \label{eq:faraday_lagrangian_l4}
\end{equation}
where $a_s$ is the $s$-wave scattering length of atoms, expressed in units of the harmonic oscillator length. The Lagrangian term that corresponds to the DDI energy is given by
\begin{equation}
    L_5(t) = - \frac{3 N a_{\mathrm{dd}}}{2} \int d{\bf r} \, d{\bf r}'\, \psi^*({\bf r}) \psi^*({\bf r}')
       U_\mathrm{dd}({\bf r - r}') \psi({\bf r}') \psi({\bf r}) \, ,
\end{equation}
where the dipolar potential reads $U_\mathrm{dd} ({\bf r}) = (1 - 3 \cos^2 \theta)/r^3$, $\theta$ is the angle between the dipoles' orientation ($z$ axis) and vector ${\bf r}$, and $a_\mathrm{dd}$ is the DDI interaction strength, that depends on the dipole moment of atoms $d$ and their mass $m$ as $a_\mathrm{dd} = \upmu_0 m d^2 / (12 \pi \hbar^2)$. Note that it is conveniently expressed in units of length and cast into a dimensionless quantity as outlined above. However, due to the spatial modulation term in the modified Gaussian ansatz,  it is not possible to perform exact integration and obtain $L_5(t)$. Using the convolution theorem, the DDI term can be written as
\begin{equation}
    L_5(t) =
        - \frac{3 N a_{\mathrm{dd}}}{2 \, (2\pi)^3} \int d{\bf k} \,
        \mathcal{F}\left[
            U_{\mathrm{dd}}
            \right]({\bf k}) \,
        \mathcal{F}\left[
            \left| \psi \right|^{2}
            \right]^2({\bf k}) \, ,\label{eq:faraday_lagrangian_l5-k}
\end{equation}
where $\mathcal{F}$ stands for the Fourier transform, and
\begin{equation}
    \mathcal{F}\left[
            \left| \psi \right|^{2}
            \right]({\bf k}) =
            B(k_x, u_x, \alpha, \beta, k) \, e^{- \frac{1}{4} ( k_x^2 u_x^2 + k_y^2 u_y^2 + k_z^2 u_z^2)} \, .
\end{equation}

The coefficient $B$ can be explicitly calculated and reads
\begin{equation}
        B(k_x, u_x, \alpha, \beta, k) = \frac{4 + 4 (e^{- \frac{k}{4} (k - 2 k_x) u_x^2} + e^{- \frac{k}{4} (k + 2 k_x) u_x^2}) \, \alpha 
                    + (2 + e^{- k (k - k_x) u_x^2} + e^{- k (k + k_x) u_x^2}) \, (\alpha^2 + \beta^2)}
            {2 \, \left[ 2 + 4 \, e^{- \frac{1}{4} k^2 u_x^2} \alpha + (1 + e^{- k^2 u_x^2 }) \, (\alpha^2 + \beta^2) \right]} \, .
\end{equation}

To proceed further, we take into account that the condensate width in the weak confinement direction is large compared to the other widths, as well as compared to the spatial period of the density waves, such that $k u_x \gg 1$. We also take into account that the wave amplitude is small immediately after the waves emerge, such that $\alpha, \beta \ll 1$. Since the integral over $\bf{k}$ in Equation~(\ref{eq:faraday_lagrangian_l5-k}) cannot be analytically performed even using these approximations, we replace $B^{2}$, stemming from the square of the Fourier transform $\mathcal{F}\left[|\psi |^2\right]$, by its average over $k_x$, and neglect all terms proportional to $e^{- k^2 u_x^2 / 8}$ and its powers, as already argued that $k u_x$ is a large quantity. The integration over $\bf{k}$ can now proceed smoothly, yielding
\begin{equation}
    L_5(t) = \frac{N a_\mathrm{dd}}{\sqrt{2 \pi} \, u_x u_y u_z} f\left( \frac{u_x}{u_z}, \frac{u_y}{u_z} \right) 
             \left( 1 - \frac{8 \alpha^2}{(2 + \alpha^2 + \beta^2)^2} \right)\, ,
    \label{eq:faraday_lagrangian_l5}
\end{equation}
where $f$ is the standard dipolar anisotropy function~\cite{PhysRevA.74.013621}.

Now that we have the explicit expression for the Lagrangian of the system $L(t)=\sum_{i=1}^5 L_i(t)$, we can derive the corresponding Euler-Lagrange equations. We assume that the wave amplitudes $\alpha$ and $\beta$ are small, such that their quadratic and higher order terms can be neglected in the equations of motion. The three equations for the phases yield $\phi_i = \dot u_i / (2 u_i)$ and can be used to eliminate the phases $\phi_i$ from the corresponding set of equations for the condensate widths $u_i$, which have the form of the second order differential equations,
\begin{equation}
    \begin{split}
    \ddot u_x + 
        \gamma^2 u_x - \frac{1}{u_x^3} - 
        \sqrt{\frac{2}{\pi}} \frac{N}{u_x^2 u_y u_z} \Bigg[
            a_s 
            - a_\mathrm{dd} f\left( \frac{u_x}{u_z}, \frac{u_y}{u_z} \right)
            + a_\mathrm{dd} \frac{u_x}{u_z} f_1\left( \frac{u_x}{u_z}, \frac{u_y}{u_z} \right)
        \Bigg]
        = 0\, ,
    \end{split}
    \label{eq:euler_lagrange_ux}
\end{equation}
\begin{equation}
    \begin{split}
    \ddot u_y + 
        \nu^2 u_y - \frac{1}{u_y^3} - 
        \sqrt{\frac{2}{\pi}} \frac{N}{u_x u_y^2 u_z} \Bigg[
            a_s 
            - a_\mathrm{dd} f\left( \frac{u_x}{u_z}, \frac{u_y}{u_z} \right)
            + a_\mathrm{dd} \frac{u_y}{u_z} f_2\left( \frac{u_x}{u_z}, \frac{u_y}{u_z} \right)
        \Bigg]
        = 0\, ,
    \end{split}
    \label{eq:euler_lagrange_uy}
\end{equation}
\begin{equation}
    \begin{split}
    \ddot u_z + 
        \lambda^2 u_z - \frac{1}{u_z^3} - 
        \sqrt{\frac{2}{\pi}} \frac{N}{u_x u_y u_z^2} \Bigg[
            a_s 
            - a_\mathrm{dd} f\left( \frac{u_x}{u_z}, \frac{u_y}{u_z} \right)
            & - a_\mathrm{dd} \frac{u_x}{u_z} f_1\left( \frac{u_x}{u_z}, \frac{u_y}{u_z} \right) \\
            & - a_\mathrm{dd} \frac{u_y}{u_z} f_2\left( \frac{u_x}{u_z}, \frac{u_y}{u_z} \right)
        \Bigg]
        = 0\, ,
    \end{split}
    \label{eq:euler_lagrange_uz}
\end{equation}
where $f_1$ and $f_2$ are partial derivatives of the anisotropy function with respect to the first and the second argument.
The Euler-Lagrange equation for the variational parameter $\beta$ yields $\beta = 2 \dot \alpha / k^2$, which we use to eliminate $\beta$ from the corresponding equation for the parameter $\alpha$, as was done with the phases. With this, the equation for $\alpha$ turns out to be the second order differential equation,
\begin{equation}
    \ddot \alpha + \left[ \frac{k^4}{4} + \sqrt{\frac{2}{\pi}} \frac{N}{u_x u_y u_z} \left( a_s 
        + a_\mathrm{dd} \, f\left( \frac{u_x}{u_z}, \frac{u_y}{u_z} \right) \right) k^2 \right] 
        \alpha = 0 \, .
    \label{eq:faraday_euler_lagrange_alpha}
\end{equation}

In the context of variational analysis of Faraday and resonant waves, the above equation of motion for the wave amplitude $\alpha$ is usually cast into the form of the Mathieu-like equation~\cite{McLachlan}
\begin{equation}
    \ddot \alpha + \left[ a(k) + \epsilon b(k) \sin 2 \tau \right]  \alpha = 0 \, .
    \label{eq:mathieu-like_form}
\end{equation}

This equation can be solved perturbatively in the small modulation amplitude $\epsilon$. Assuming a solution in the form of a harmonic oscillator
\begin{equation}
    \alpha(\tau, \epsilon) = P(\epsilon \tau) \cos \left( \tau \sqrt{a(k)} \right) + Q(\epsilon \tau) \sin \left( \tau \sqrt{a(k)} \right) \, ,\label{eq:mathieu-like_solution}
\end{equation} 
we obtain that functions $P$ and $Q$ are exponentials of the form $e^{\pm i \xi \epsilon\tau}$, where $\xi$ is a complex number. The existence of the imaginary part of $\xi$ leads to the instability, i.e., to the exponential growth of the wave amplitude, which yields Faraday or resonant waves. It was shown in Reference~\cite{McLachlan} that the nonvanishing imaginary part of $\xi$ appears for $a(k) = n^2$, where $n \in \mathbb{N}$, and this represents the mathematical form of the instability condition.

   To cast Equation (\ref{eq:faraday_euler_lagrange_alpha}) into the Mathieu-like form (Equation \eqref{eq:mathieu-like_form}), we need to take into account that the radial trap frequencies are modulated, such that the corresponding trap aspect ratio is given by $\nu(t)=\lambda(t) = \lambda_0 (1 + \epsilon \sin \eta_m t)$, where $\lambda_0=\Omega_0/\omega_r$. This generates the dynamics of the system and we need to obtain approximate expressions for the condensate widths in order to get explicit form of the quantities $a(k)$ and $b(k)$. We assume that the condensate width $u_x$ slowly varies, and can be taken to be constant at the onset of instability. We also assume that second derivatives of the radial widths $u_y$ and $u_z$, with respect to time, can be neglected, since they are proportional to the small modulation amplitude $\epsilon$. Furthermore, for simplicity, we assume $u_y \approx u_z \equiv u_\rho$, which now satisfies the modified Equation (\ref{eq:euler_lagrange_uy}) or (\ref{eq:euler_lagrange_uz}) in the form
\begin{equation}
    \lambda^2(t) u_\rho^4 = 1 + \sqrt{\frac{2}{\pi}} \frac{N}{u_x} \left[
            a_s + \frac{a_\mathrm{dd}}{2} f_s \left( \frac{u_\rho}{u_x} \right) 
            - a_\mathrm{dd} f_s^{\prime}\left( \frac{u_\rho}{u_x} \right)
        \right] \, ,
    \label{eq:cylindrical_xaxis_urho}
\end{equation}
where $f_s(x)=f(x,x)$. On the right-hand side of the above equation, we assume that the ratio $u_\rho / u_x$ is constant and equal to the corresponding ratio for the ground state, which can be calculated from Equations~(\ref{eq:euler_lagrange_ux})--(\ref{eq:euler_lagrange_uz}). If we express $u_\rho^2$ from Equation (\ref{eq:cylindrical_xaxis_urho}), and use it to estimate the quantity $u_y u_z \approx u_\rho^2$ in Equation (\ref{eq:faraday_euler_lagrange_alpha}), we obtain
the equation for the variational parameter $\alpha$ in the form
\begin{equation}
    \ddot \alpha + \left[ \frac{k^4}{4} + \frac{\Lambda k^2}{4} \lambda(t) \right]  \alpha = 0 \, ,\label{eq:almost_mathieu}
\end{equation}
where $\Lambda$ is given by
\begin{equation}
    \Lambda = \frac{4 \sqrt{\frac{2}{\pi}} \, N \left[ a_s 
        - \frac{a_\mathrm{dd}}{2} \, f_s \left( \frac{u_\rho}{u_x} \right) \right]}
        {u_x \left\{ 1 + \sqrt{\frac{2}{\pi}} \frac{N}{u_x} \left[
            a_s + \frac{a_\mathrm{dd}}{2} f_s \left( \frac{u_\rho}{u_x} \right) 
            - a_\mathrm{dd} f_s^{\prime}\left( \frac{u_\rho}{u_x} \right)
        \right] \right\}^{1/2} } \, .
\end{equation}

After inserting the explicit form for $\lambda(t)$ into Equation~(\ref{eq:almost_mathieu}), we still need to make a variable change $\eta_m t \rightarrow 2 \tau$ in order to transform it into the Mathieu-like Equation (\ref{eq:mathieu-like_form}). This finally yields the expressions for the coefficients $a(k)$ and $b(k)$,
\begin{equation}
    a(k) =  \frac{k^4}{\eta_m^2} + \frac{\lambda_0 \Lambda k^2}{\eta_m^2} \, , \qquad b(k) = \frac{\lambda_0 \Lambda k^2}{\eta_m^2} \, .
\end{equation}

As previously discussed, the instability condition for the Faraday waves reads $a(k) = 1$, which can be used to calculate the wave vector of density waves shortly after their emergence,
\begin{equation}
    k_{F} = \sqrt{- \frac{\lambda_0 \Lambda}{2} + \sqrt{\frac{\lambda_0^2 \Lambda^2}{4} + \eta_m^2}} \, .
    \label{eq:k_F}
\end{equation}

This represents our analytical result for the wave vector $k_F$ and the spatial period $\ell_F = 2 \pi / k_F$ of the Faraday waves, which can be directly compared with numerical or experimental results. Let us also stress that the above analysis is consistent with the main characteristic of the Faraday waves, namely, that their oscillation frequency is half that of the driving frequency. This can be concluded according to $\tau = \eta_m t / 2$ and Equation (\ref{eq:mathieu-like_solution}), where we see that indeed the solution of the derived Mathiue-like equation oscillates with the frequency whose aspect ratio is $\eta_m / 2$, i.e., with the frequency $\omega_m/2$.

If the modulation frequency is close to one of the characteristic oscillation modes of the system, it~will exhibit resonant behavior, which is suppressed for an arbitrary value of the modulation frequency. While the system's dynamics will certainly include the Faraday mode at the frequency $\omega_m / 2$ even close to a resonance, the resonant mode with the frequency $\omega_m$ will have a larger amplitude and will develop much faster. Although it is clear that the above analysis would break down, the condition for the emergence of resonant waves still corresponds to $a(k) = 2^2$, i.e., the wave vector of the resonant wave is given by

\begin{equation}
    k_{R} = \sqrt{- \frac{\lambda_0 \Lambda}{2} + \sqrt{\frac{\lambda_0^2 \Lambda^2}{4} + 4 \eta_m^2}} \, .
    \label{eq:k_R}
\end{equation}

In that case, according to $\tau = \eta_m t / 2$ and Equation (\ref{eq:mathieu-like_solution}), the resonant density wave will oscillate with the frequency whose aspect ratio is $(\eta_m/2)\sqrt{2^2}=\eta_m$, i.e., with the frequency $\omega_m$. Depending on the system's parameters, higher resonant modes can also appear corresponding to the conditions $a(k) = n^2$, where $n$ is an integer, corresponding to the oscillation frequencies $n\omega_m/2$. 

\section{Faraday Waves in Chromium, Erbium, and Dysprosium Condensates}
\label{sec:faraday}

   To study Faraday waves in dipolar condensates, we   performed extensive numerical simulations of the real-time dynamics and solved the dipolar Gross-Pitaevskii equation using the programs described in References~\cite{ComputPhysCommun.180.1888,ComputPhysCommun.183.2021,ComputPhysCommun.195.117,ComputPhysCommun.200.406,ComputPhysCommun.200.411,ComputPhysCommun.204.209,ComputPhysCommun.220.503,ComputPhysCommun.209.190,ComputPhysCommun.240.74}. The parameters of these simulations match the physical parameters of BECs of chromium $^{52}$Cr, erbium $^{168}$Er, and dysprosium $^{164}$Dy, which, respectively ,have the dipole moments $d=6\upmu_\mathrm{B}$, $d=7\upmu_\mathrm{B}$, and $d=10\upmu_\mathrm{B}$, where $\mathrm{B}$ is the Bohr magneton. The~corresponding background $s$-wave scattering lengths are $a_s=105 a_0$, $a_s=100 a_0$, and $a_s=100 a_0$, where $a_0$ is the Bohr radius. We used these interaction strengths, unless otherwise specified.

\begin{figure}[H]
    \centering
        \includegraphics[width=0.76\textwidth]{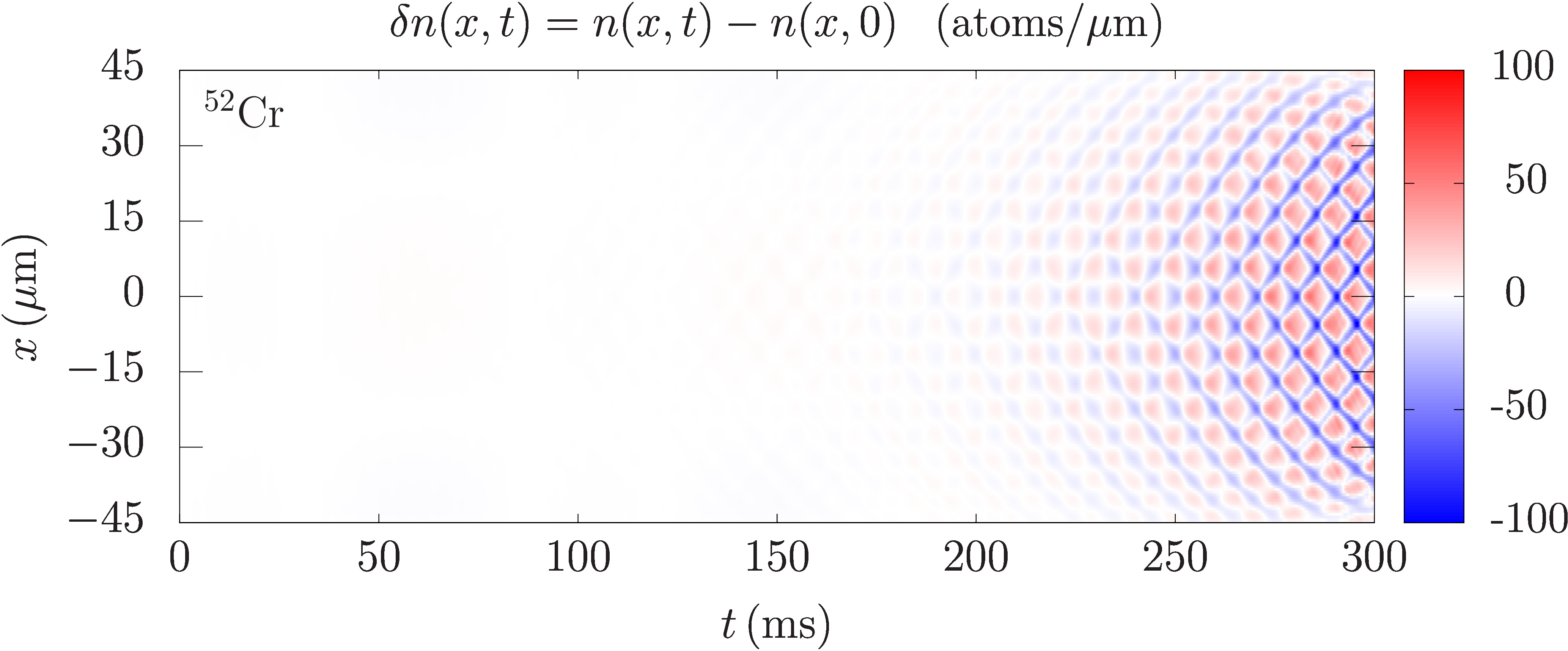}
        \includegraphics[width=0.76\textwidth]{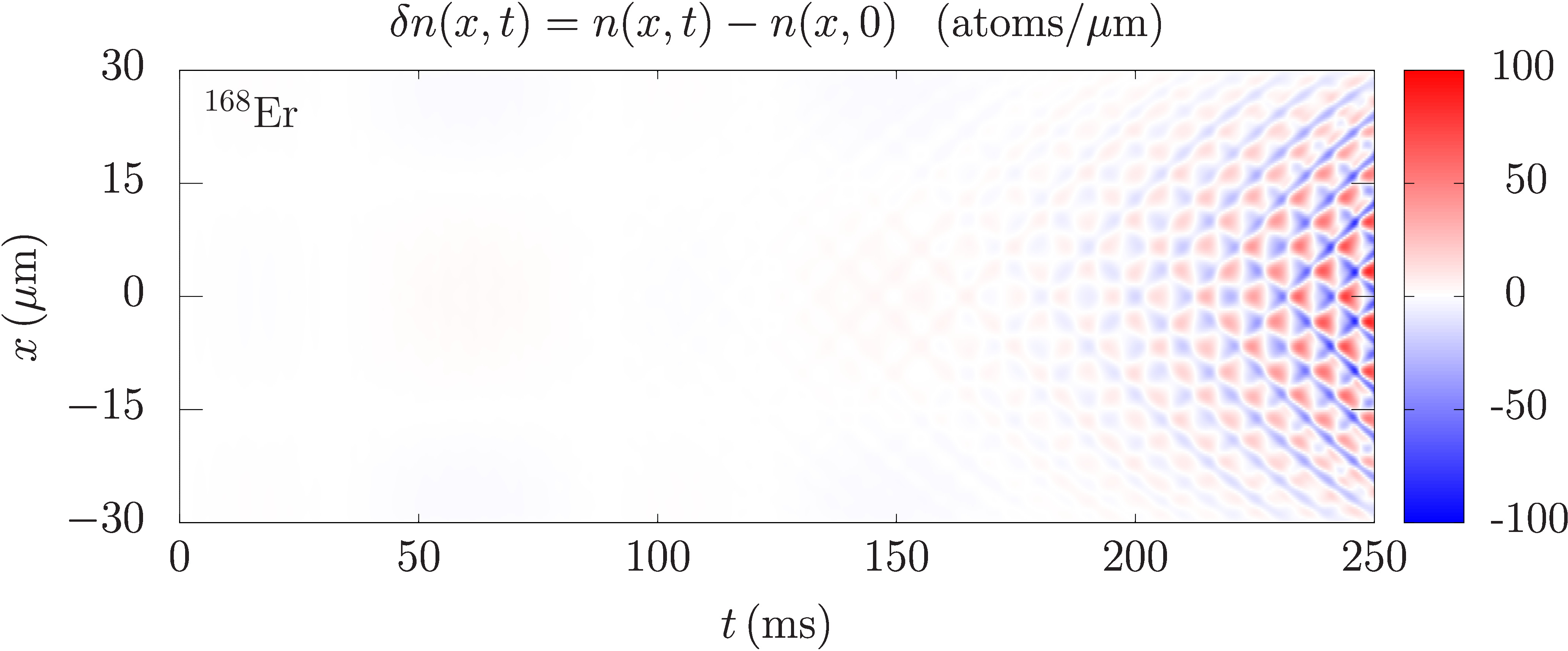}
        \includegraphics[width=0.76\textwidth]{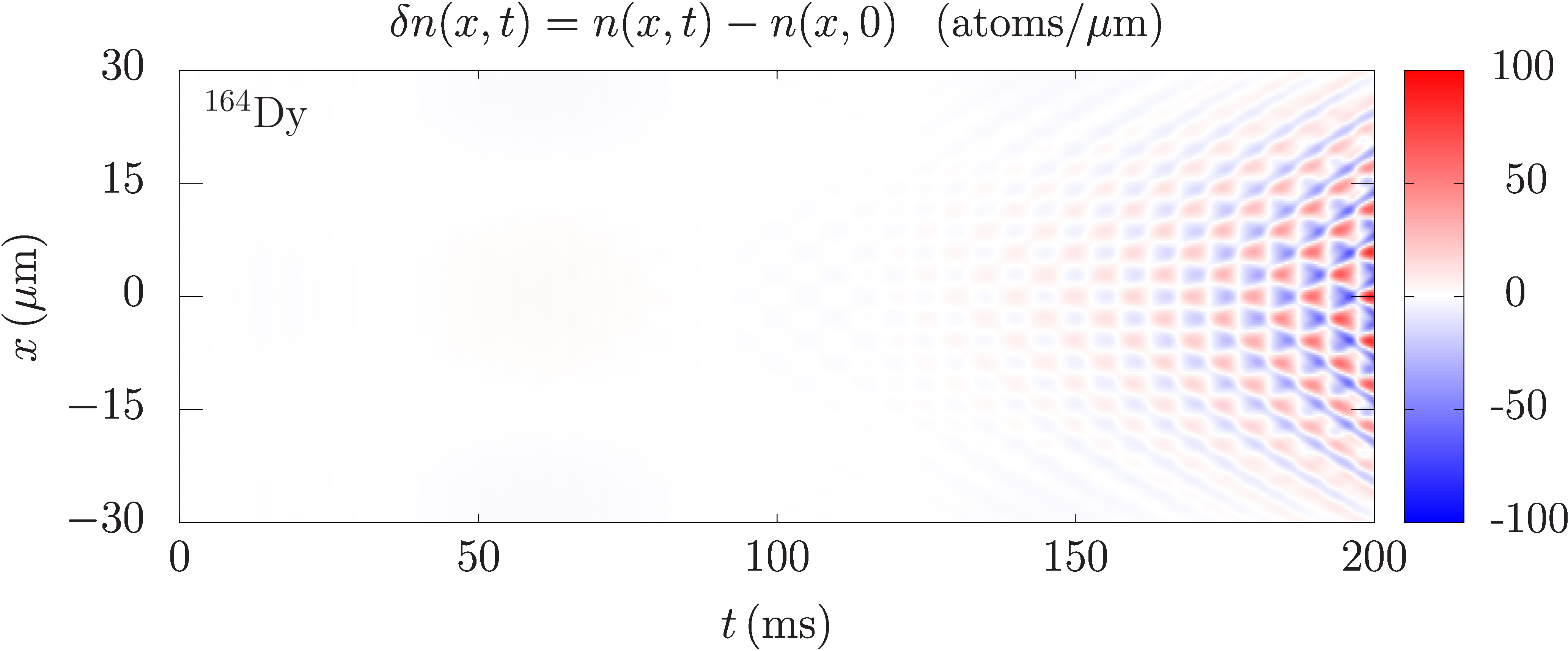}
    \caption{Time evolution of the integrated density profile variation $\delta n(x, t)$ in the weak-confinement direction for a BEC of chromium $^{52}$Cr (\textbf{top}),
     erbium $^{168}$Er (\textbf{middle}),
      and dysprosium $^{164}$Dy (\textbf{bottom}),
       for the modulation frequency $\omega_m = 200 \times 2 \pi$~Hz and amplitude $\epsilon = 0.2$, and the system parameters given in Section~\ref{sec:variational}.}
    \label{fig:faraday_ndenx_crerdy}
\end{figure}

As discussed previously, Faraday waves are expected as a main excitation mode of the system when the modulation frequency $\omega_m$ does not match any of the characteristic frequencies of the system. For this reason, we used the value $\omega_m = 200 \times 2 \pi$~Hz, for which we   verified that these conditions are satisfied.    To characterize the density waves, we typically analyze their FFT spectra in the time-frequency and     spatial-frequency domains. However, instead of directly analyzing their density profiles, for FFT, it is advantageous to have a clearer signal, which can be obtained by considering only the density variations compared to the initial state, i.e., the ground state of the system, before the modulation is switched on. Therefore, Figure~\ref{fig:faraday_ndenx_crerdy} shows time dependence of the integrated density profile variations in the weak confinement direction $\delta n(x, t)=n(x, t)-n(x, t=0)$. Here, $n(x, t)$ is the column density profile calculated by integrating the 3D condensate density $|\psi|^2$ over the radial coordinates $y$ and $z$. 

The emergence of spatial patterns is clearly visible for all three atomic species after around 150~ms. This is consistent with earlier experimental observations~\cite{PhysRevLett.98.095301,PhysRevA.81.053627,PhysRevX.9.011052} and theoretical results~\cite{PhysRevA.76.063609,PhysRevA.85.023613,PhysRevA.89.023609}. The~density waves in $x$ direction from Figure~\ref{fig:faraday_ndenx_crerdy} take time to develop and are a result of the transfer of energy from the modes that are directly excited in the radial directions, where the trap is modulated. On the other hand, the density waves in the radial directions (which are not shown here) emerge immediately after the modulation is switched on at $t = 0$, and their frequency is equal to the modulation frequency. By looking at Figure~\ref{fig:faraday_ndenx_crerdy}, we can even estimate the main oscillation frequency, e.g., counting the number of maxima or minima in a given time interval. For instance, in the last 50~ms in each of the panels in Figure~\ref{fig:faraday_ndenx_crerdy}, we count  five periods, which corresponds to the frequency $100 \times 2 \pi~\mathrm{Hz} = \omega_m / 2$. This~is a distinguishing characteristic of Faraday waves, and therefore we can directly determine that in this case the system develops this type of collective oscillations.

\begin{figure}[H]
    \centering
        \includegraphics[width=0.31\textwidth]{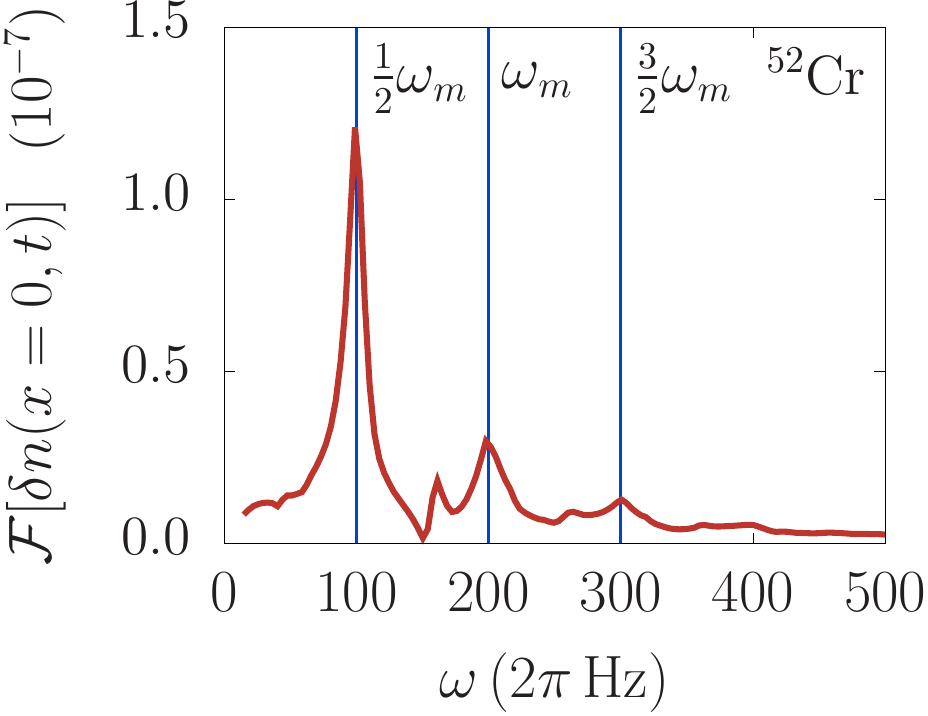}
        \includegraphics[width=0.31\textwidth]{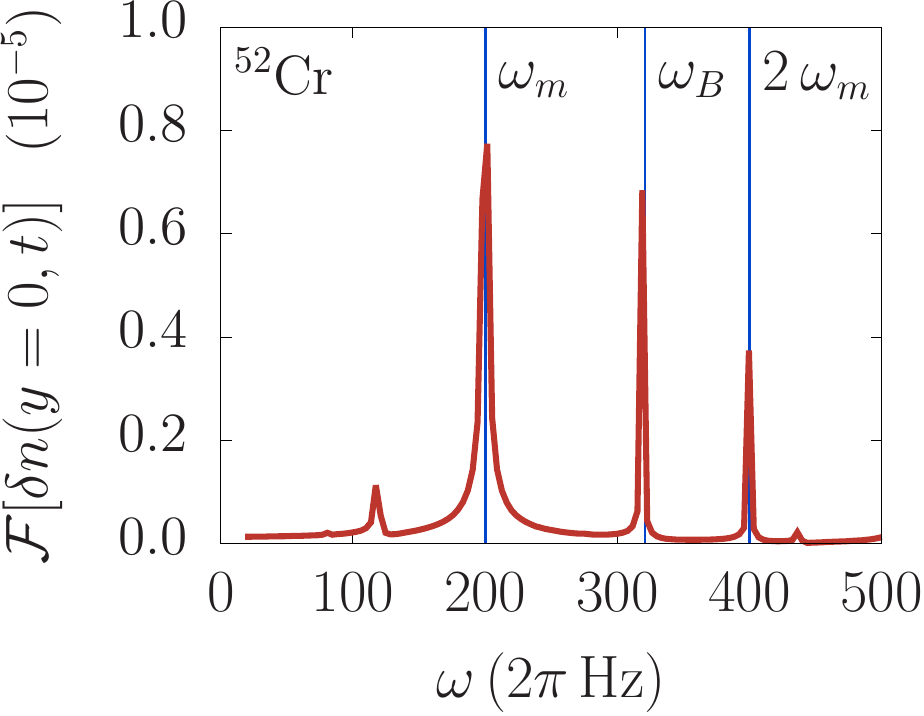}
        \includegraphics[width=0.31\textwidth]{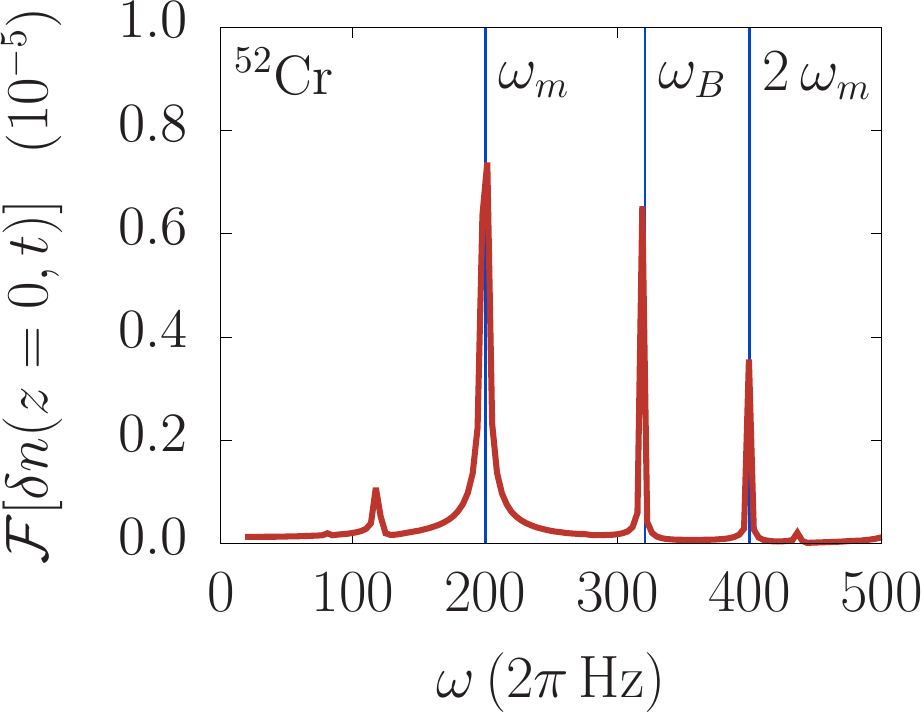}
        \includegraphics[width=0.31\textwidth]{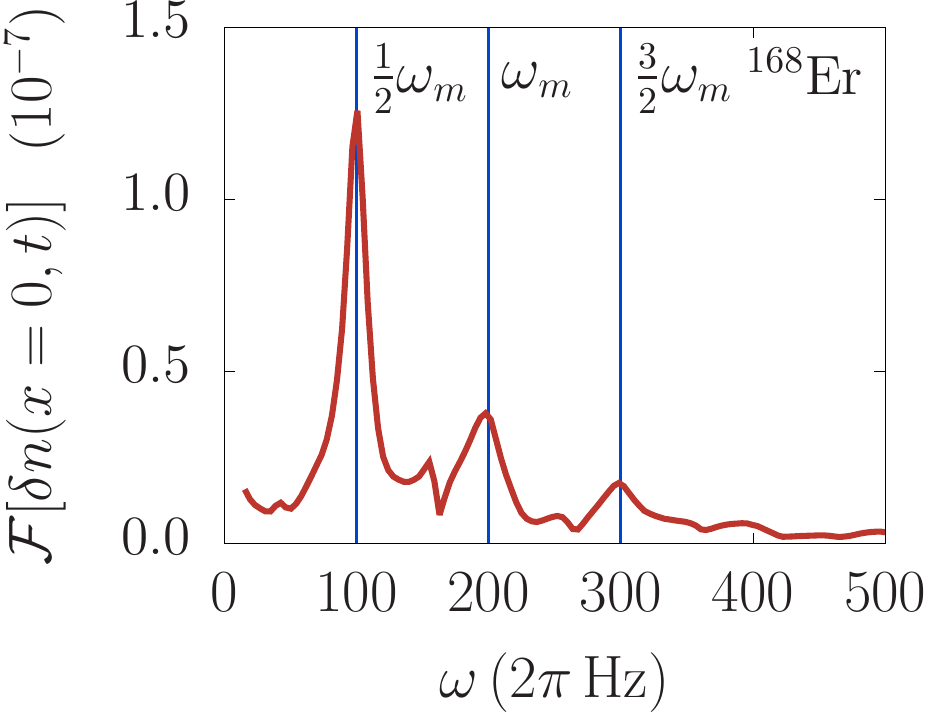}
        \includegraphics[width=0.31\textwidth]{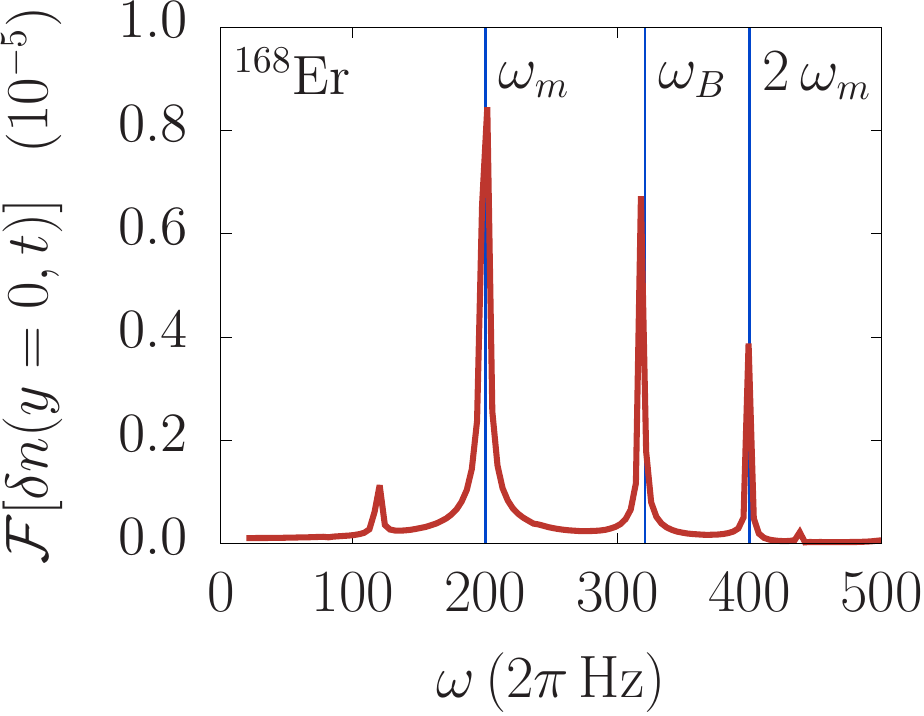}
        \includegraphics[width=0.31\textwidth]{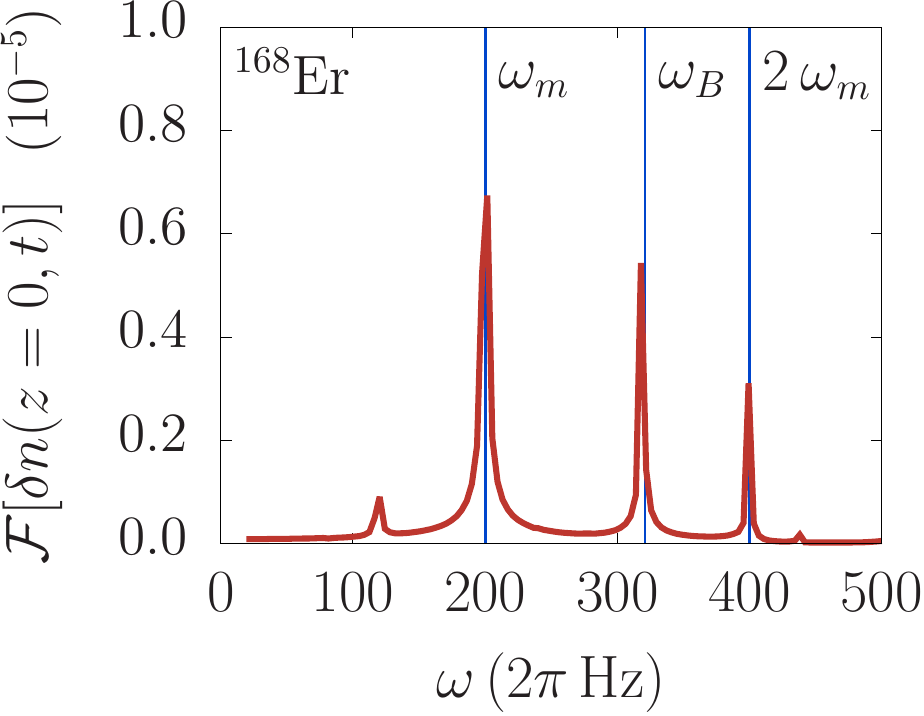}
        \includegraphics[width=0.31\textwidth]{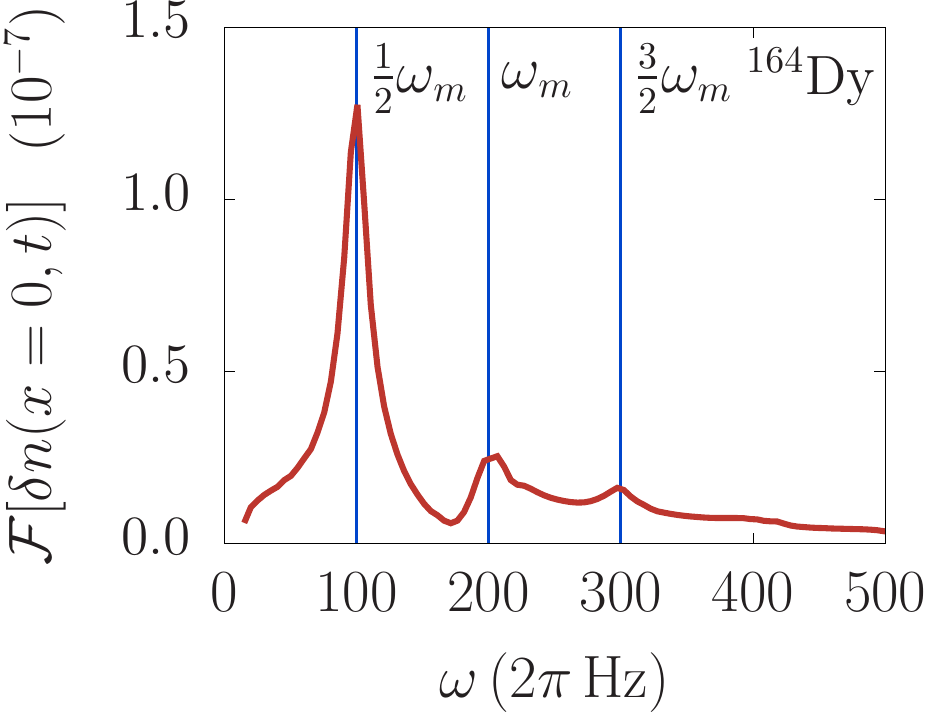}
        \includegraphics[width=0.31\textwidth]{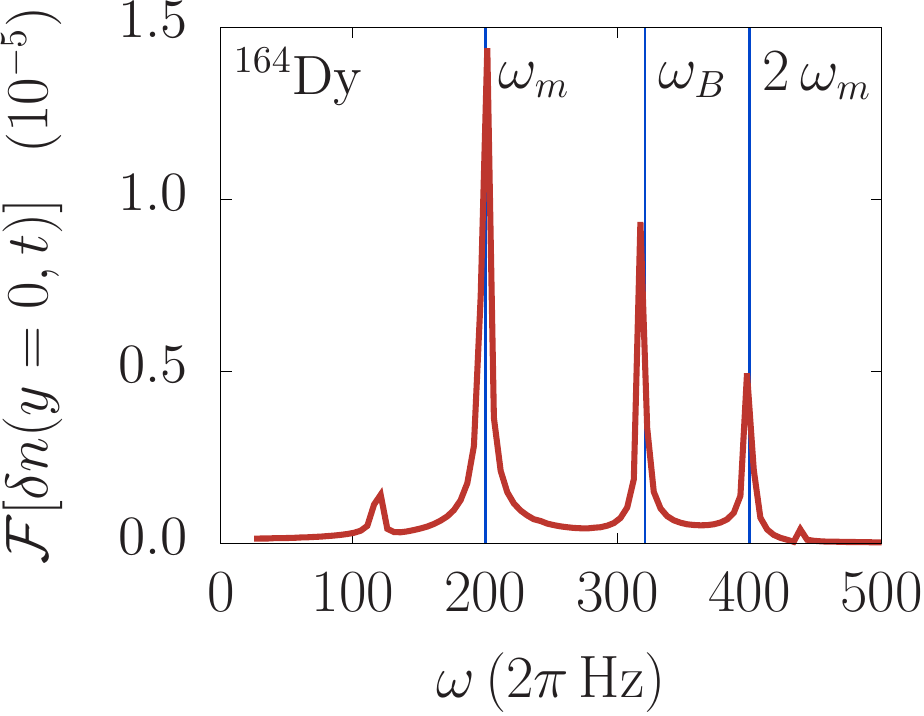}
        \includegraphics[width=0.31\textwidth]{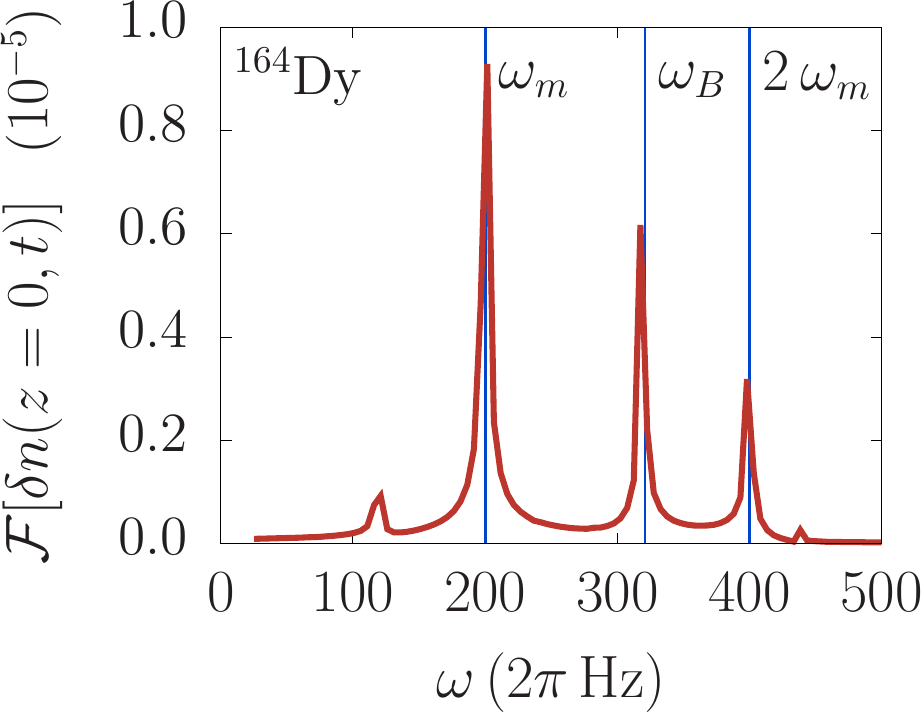}
    \caption{The Fourier spectrum in the time-frequency domain of the integrated 1D density profile variations of Faraday waves at the trap center $\delta n(x = 0, t)$ in $x$ direction (\textbf{left column}),
     $\delta n(y = 0, t)$ in $y$ direction (\textbf{middle column}),
      and $\delta n(z = 0, t)$ in $z$ direction (\textbf{right column})
       for a BEC of chromium $^{52}$Cr (\textbf{top row}),
        erbium $^{168}$Er (\textbf{middle row}),
         and dysprosium $^{164}$Dy (\textbf{bottom row}).
         Vertical blue lines represent theoretical predictions, where $\omega_m / 2$ corresponds to Faraday waves, $\omega_m$, $3\omega_m/2$, and $2 \, \omega_m$ to resonant waves, and $\omega_B$ is the variational result for the breathing mode frequency, which is obtained by linearization of the equations of motion from Section~\ref{sec:variational}.}
    \label{fig:faraday_fft_t}
\end{figure}

However, this way we can determine only the main excitation modes. The dynamics of the system contains other modes as well, and over the time they can develop and even start to dominate the behavior of the system. Therefore, it is important to analyze the spectra in more detail. This is done in Figure~\ref{fig:faraday_fft_t} for all integrated density profile variations, separately for each spatial direction. For simplicity, the FFT analysis is performed for the profiles at the trap center. As expected, in the weak confinement direction  (left column of Figure~\ref{fig:faraday_fft_t}), the main excitation mode has a frequency $\omega_m / 2$. In addition to this, we observe two other modes, at $\omega_m$ and $3 \, \omega_m / 2$. This is expected from the theoretical analysis in Section \ref{sec:variational}, but could not be discerned directly from the density profiles or their variations.

In the Fourier spectra of the integrated density profile variations in the radial directions (middle and right columns of Figure~\ref{fig:faraday_fft_t}), we see a somewhat richer set of excitation modes. In addition to the main mode corresponding to the trap modulation at $\omega_m$, we see that also the breathing mode is excited at the frequency $\omega_B \approx 321 \times 2 \pi$~Hz. This value can be calculated by linearizing the equations of motion from Section~\ref{sec:variational}. The spectra prominently contain the second modulation harmonic at $2 \, \omega_m$ as well. We also see some other peaks, for instance the small peak at around $120 \times 2 \pi$~Hz, which can be due to the linear combination of the modes $\omega_B - \omega_m$. However, such an identification would require further theoretical and numerical analysis, which is out of the scope of the present paper.

\begin{figure}[H]
    \centering
        \includegraphics[width=0.31\textwidth]{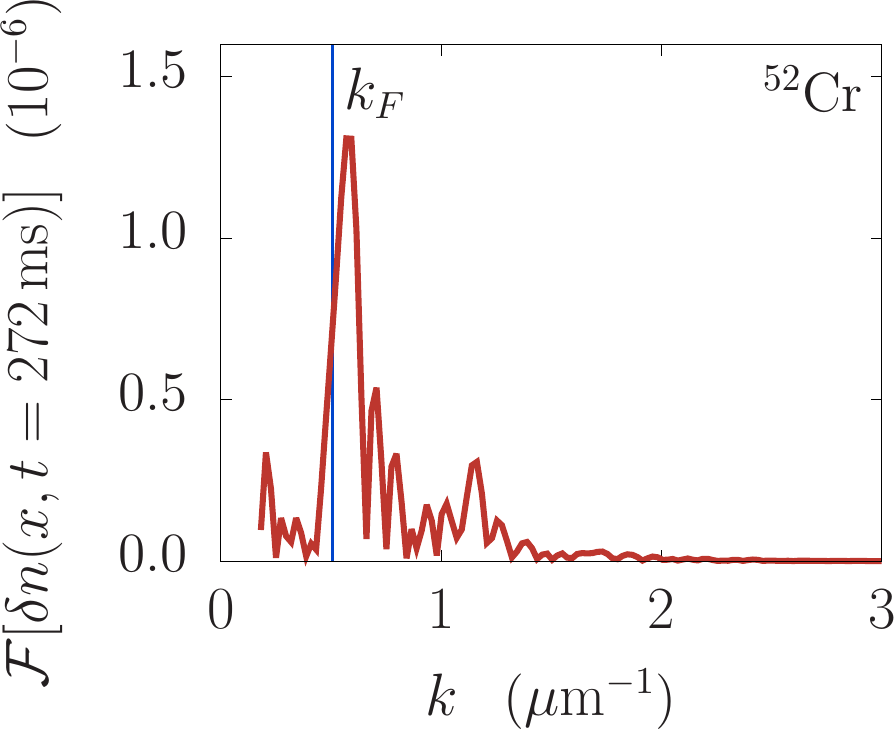}
        \includegraphics[width=0.31\textwidth]{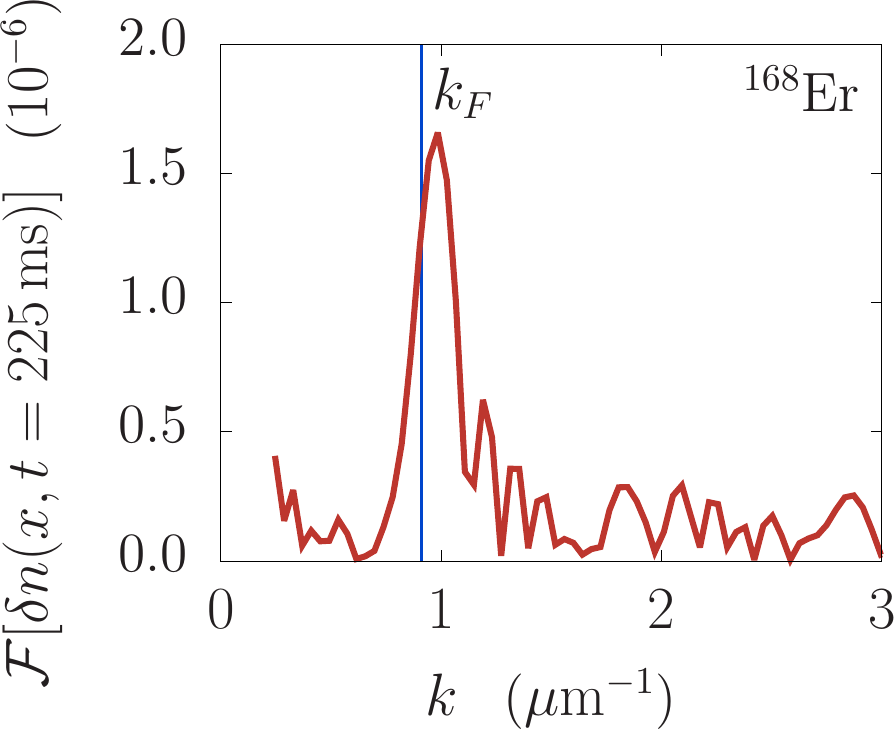}
        \includegraphics[width=0.31\textwidth]{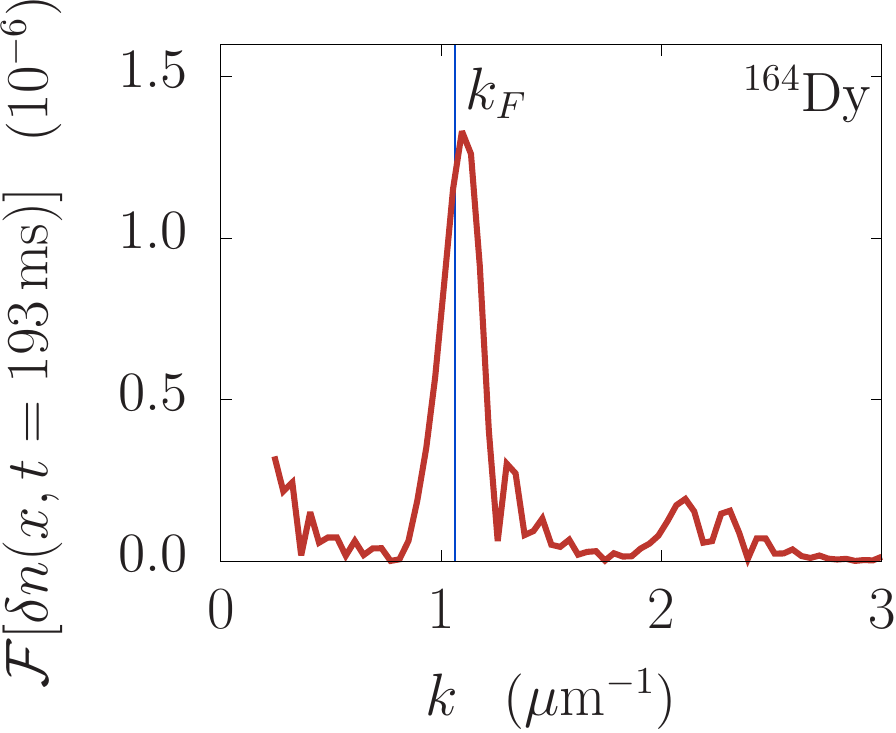}
    \caption{The Fourier spectrum in the spatial-frequency domain of the integrated 1D density profile variations of Faraday waves in $x$ direction $\delta n(x, t = 272~\mathrm{ms})$ for $^{52}$Cr (\textbf{left}),
     $\delta n(x, t = 225~\mathrm{ms})$ for $^{168}$Er (\textbf{middle}),
      and $\delta n(x, t = 193~\mathrm{ms})$ for $^{164}$Dy (\textbf{right})
       BECs with $N = 10^{4}$ atoms. The corresponding density profile variations are shown in Figure~\ref{fig:faraday_ndenx_crerdy}. Vertical blue lines represent theoretical predictions for the wave vector $k_F$ of the Faraday waves, Equation (\ref{eq:k_F}).}
    \label{fig:faraday_fft_k}
\end{figure}

While the Fourier analysis in the time-frequency domain can be used to determine the character of the induced density waves (Faraday, collective, and resonant), the analysis in the spatial-frequency domain enables us to characterize the density patterns and calculate their spatial period. This is illustrated in Figure~\ref{fig:faraday_fft_k} for Faraday waves for all three considered atomic species. The integrated density profile variations are analyzed at appropriate times, which are determined to correspond to the evolution stage when Faraday waves have fully emerged, but the system is still far from the violent dynamics that inevitably follows after the long driving period.

In all three panels of Figure~\ref{fig:faraday_fft_k}, the main peak corresponds to the wave vector $k_F$ of the Faraday waves, and we see significant differences: for $^{52}$Cr, we obtained $k_F = 0.57 \, \upmu \mathrm{m}^{-1}$, yielding the spatial period $\ell_F = 2 \pi / k_F = 11.02 \, \upmu \mathrm{m}$; for $^{168}$Er, we obtained $k_F = 0.98 \, \upmu \mathrm{m}^{-1}$ and $\ell_F = 6.41 \,\upmu \mathrm{m}$; and, for $^{164}$Dy, we obtained $k_F = 1.10 \, \upmu \mathrm{m}^{-1}$ and $\ell_F = 5.71 \,\upmu \mathrm{m}$. The variational analysis presented in Section~\ref{sec:variational} yields results which are in good agreement with the numerical ones, namely $k_F = 0.51 \, \upmu \mathrm{m}^{-1}$ for $^{52}$Cr, $k_F = 0.91 \, \upmu \mathrm{m}^{-1}$ for~\mbox{$^{168}$Er}, and $k_F = 1.06 \, \upmu \mathrm{m}^{-1}$ for $^{164}$Dy. These variational results are shown in Figure~\ref{fig:faraday_fft_k} by vertical blue lines, which illustrate their agreement with the Fourier analysis. The presented spectra also contain some additional peaks that correspond to other geometrical features of the analyzed density profile variations, such as the condensate widths and their higher harmonics, as well as the higher harmonics of the Faraday waves periods, and linear combinations of all of these. However, they are not of interest for our analysis and we did not study them further.

Note that the spatial period of Faraday waves can   also be      determined by directly looking at the density profile variations in Figure~\ref{fig:faraday_ndenx_crerdy}, and estimating the spacing between the consecutive minima or maxima at the appropriate evolution time. For instance, for chromium, we count three minima over the spatial extent of 30~$\upmu$m, yielding an estimate $\ell_F \approx 10 \upmu \mathrm{m}$, and similarly for other species. Obviously, such estimates are not as precise as the Fourier analysis results, and therefore we rely on FFT spectra to systematically determine the spatial periods of Faraday waves and their functional dependencies on the contact and the DDI strength.

\section{Interaction Effects and Properties of Faraday Waves}
\label{sec:interaction}

In the previous section, we   show  how the Fourier analysis can be used to calculate the spatial period of Faraday waves. Next, we systematically studied the interaction effects, namely how the contact and the DDI strength affect the properties of generated density waves. First, we explored the influence of the contact interaction on the emergence time and the spatial period of Faraday waves for a fixed value of the DDI strength. In experiments, this can be achieved by employing the Feshbach resonance technique, which allows   tuning $a_s$ by changing the external magnetic field, thus changing the electronic structure of atoms and their scattering properties.

The existence of Faraday waves is a consequence of nonlinearity of the system, i.e., the presence of the contact and the DDI terms in the Hamiltonian. In a linear system, described by the pure Schr\" odinger equation, the harmonic modulation of the trap in the radial direction would not be transferred into the longitudinal direction. Therefore, the emergence time of Faraday waves (and other types of density waves in the longitudinal direction) critically depends on the strength of interatomic interactions. However, if interaction strengths become sufficiently large, the emergence time is less sensitive to their changes. Since the DDI is strong in erbium and dysprosium, we can expect that the emergence time of Faraday waves significantly depends on the contact interaction strength only in chromium, where $a_\mathrm{dd}$ is small. 

\begin{figure}[H]
    \centering
        \includegraphics[width=0.76\textwidth]{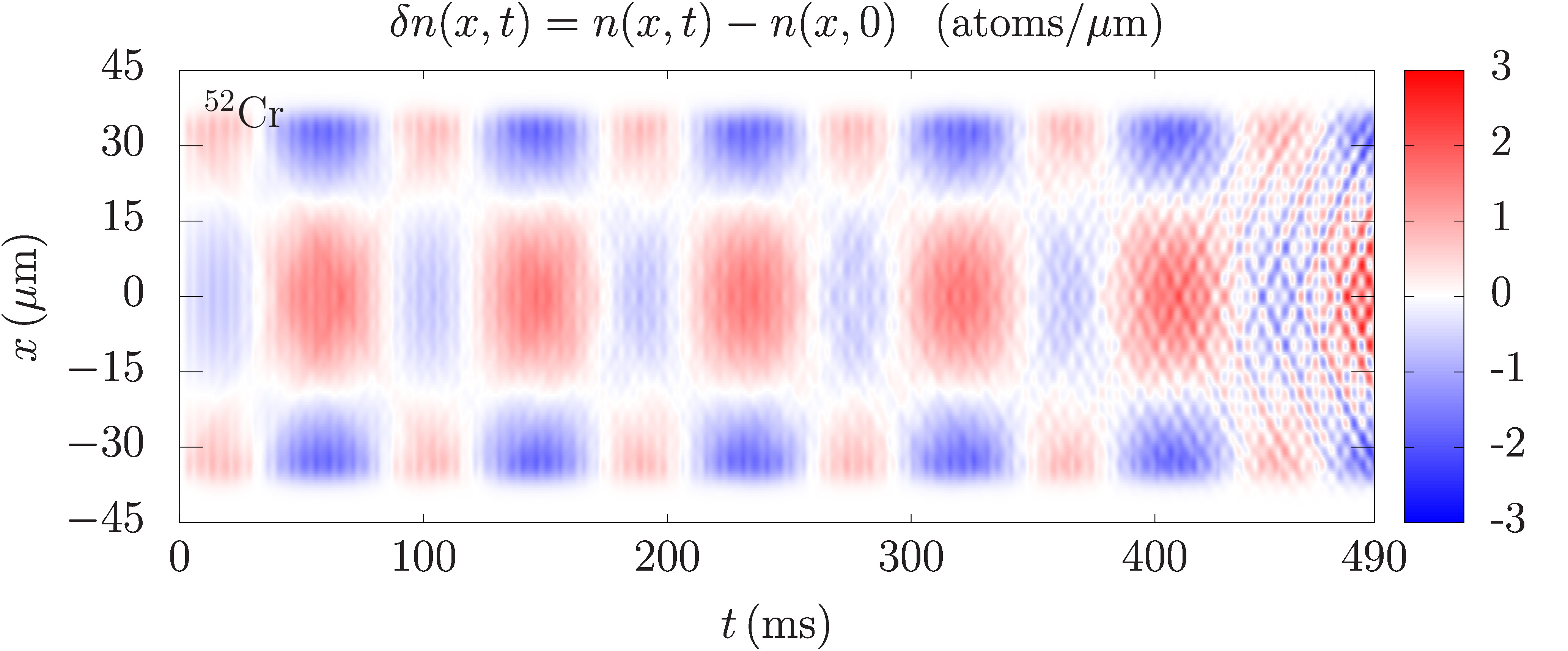}
        \includegraphics[width=0.76\textwidth]{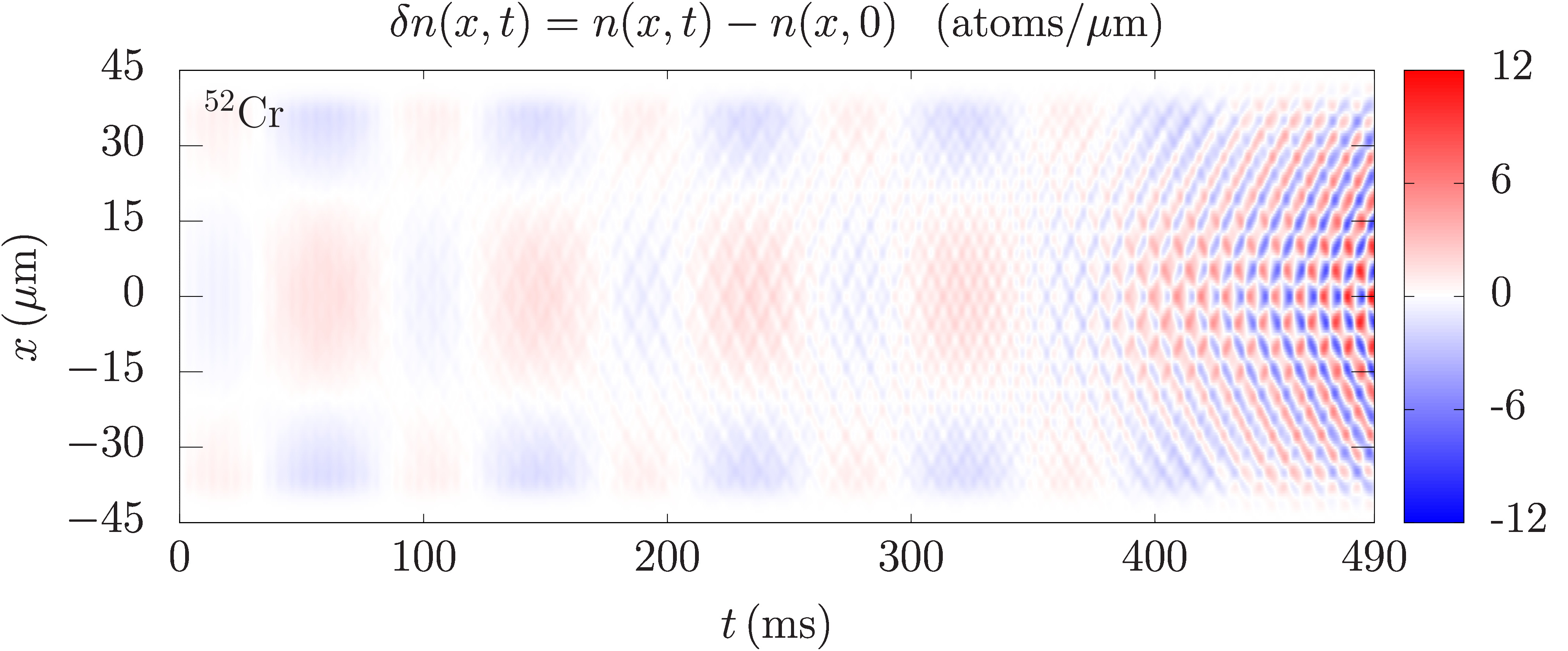}
        \includegraphics[width=0.76\textwidth]{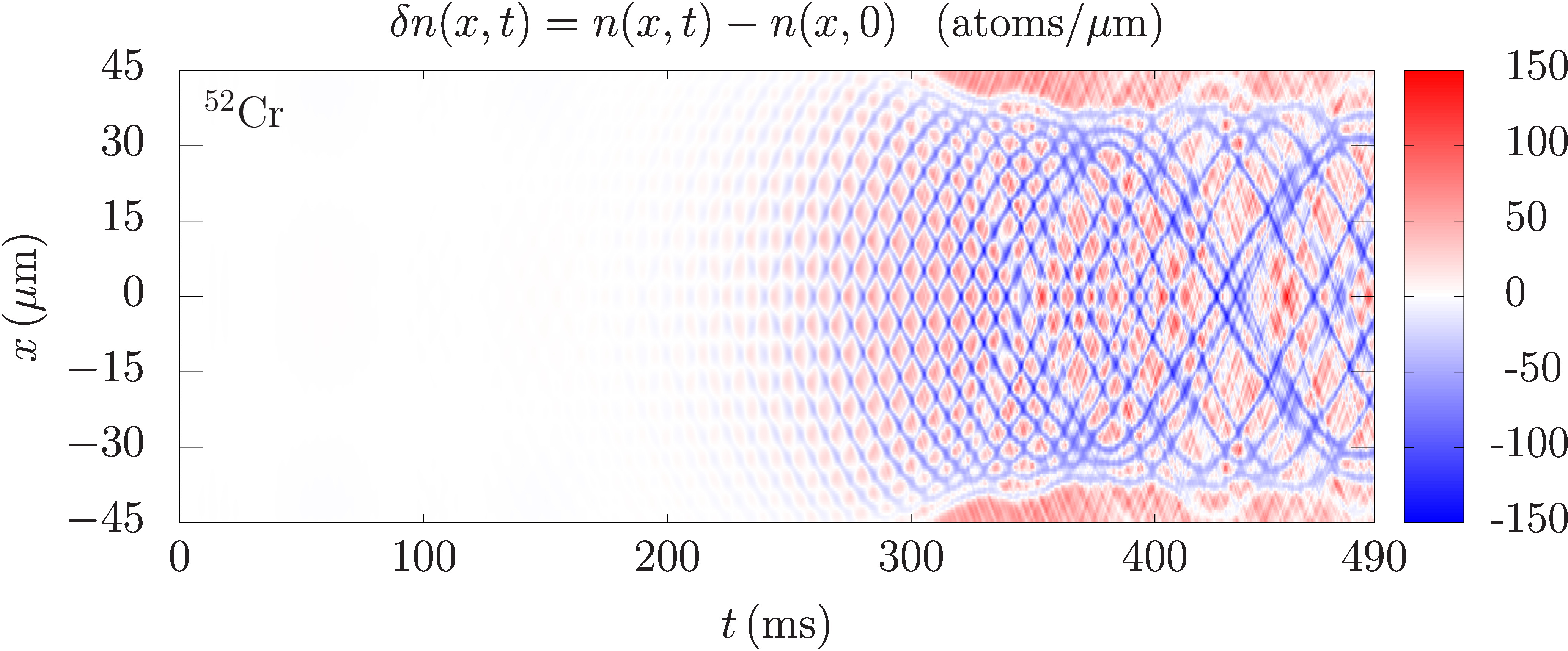}
    \caption{Emergence of Faraday waves for different strengths of the contact interaction: $a_s = 60 \, a_0$ (\textbf{top}),
     $a_s = 80 \, a_0$ (\textbf{middle}),
      and $a_s = 150 \, a_0$ (\textbf{bottom})
       for a BEC of $^{52}$Cr. We observe that Faraday waves emerge faster as the contact interaction strength increases.}
    \label{fig:faraday_time}
\end{figure}

This is illustrated in Figure~\ref{fig:faraday_time}, where we see the density profile variations for chromium for three different values of $a_s$. Let us first note that the amplitude of density variations is much smaller in the top panel for $a_s = 60 \, a_0$ than in the middle panel for $a_s = 80 \, a_0$, and significantly smaller than in the bottom panel for $a_s = 150 \, a_0$. This is also evident from the fact that in the top and middle panel we can clearly see the quadrupole collective oscillation mode, which has a frequency of around $\omega_Q = 12 \times 2 \pi$~Hz. This can be estimated from the figure and compared to the variational value of $\omega_Q$, which can be obtained by linearizing the equations of motion in Section~\ref{sec:variational}. When the interaction is sufficiently large, the amplitude of Faraday waves is much larger than those of the collective modes, and they cannot be even discerned in the bottom panel in Figure~\ref{fig:faraday_time}. Only for a weak interaction the amplitude of the Faraday waves is comparable to the amplitude of the collective modes, and this is     why we can see them all for small values of $a_s$.

As with all other excitations, Faraday waves start to develop immediately after the modulation is switched on. The question on their emergence time is related to their amplitude, which is time-dependent and grows exponentially, as can be seen from the solution  (Equation \eqref{eq:mathieu-like_solution}) of the Mathieu-like equation that describes the dynamics of the Faraday density oscillations. The imaginary part of the parameter $\xi$ in Equation (\ref{eq:mathieu-like_solution}) is responsible for the exponential growth of the Faraday waves' amplitude, which is not the case for collective modes. Therefore, in practical terms, the definition of the emergence time of Faraday waves is always arbitrary and can be expressed as a time needed for the density variations to reach a certain absolute or relative (compared to the total density) value. One can even relate this to the experimental point of view, where there is a threshold for the density variations that can be observed, due to measurement errors. However, in numerical simulations, there are no such limitations and one can easily use an arbitrary definition to estimate the emergence time of density waves. The more relevant quantity to study would be the exponent that governs the growth of the wave amplitude, which depends on the interaction strength.

\begin{figure}[H]
    \centering
        \includegraphics[width=0.31\textwidth]{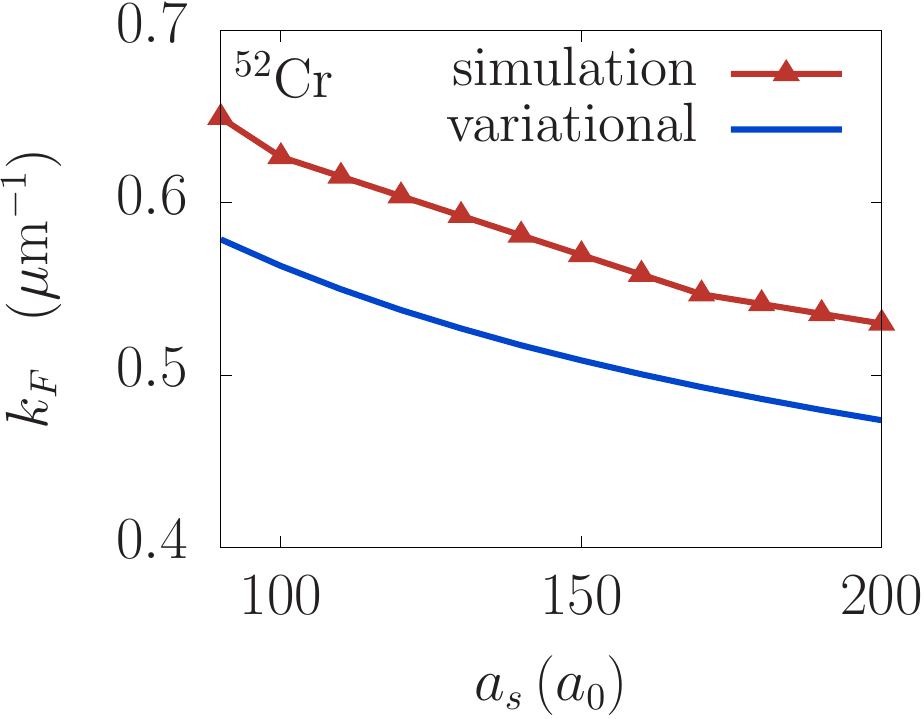}
        \includegraphics[width=0.31\textwidth]{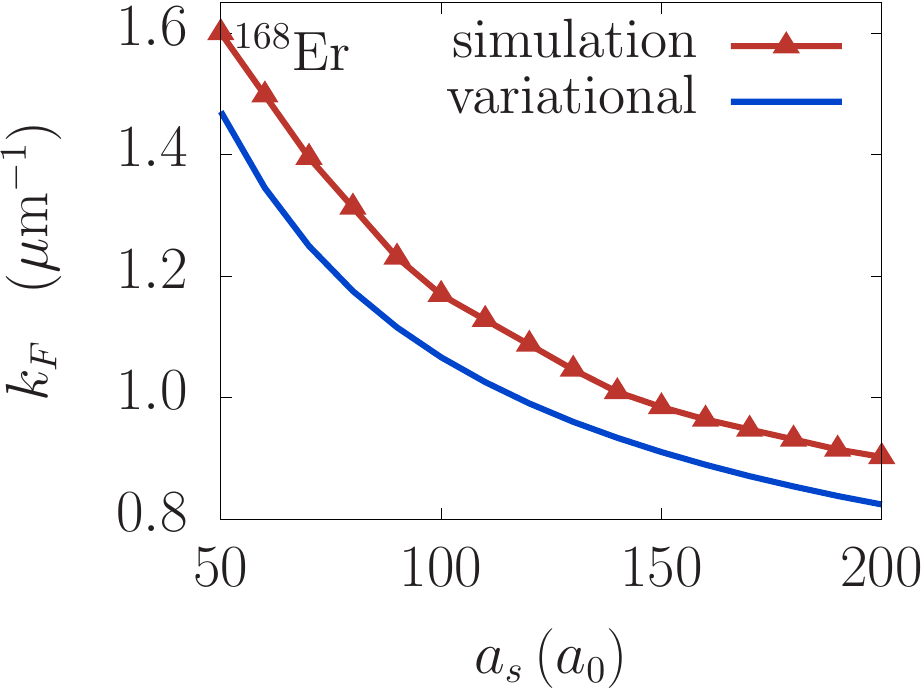}
        \includegraphics[width=0.31\textwidth]{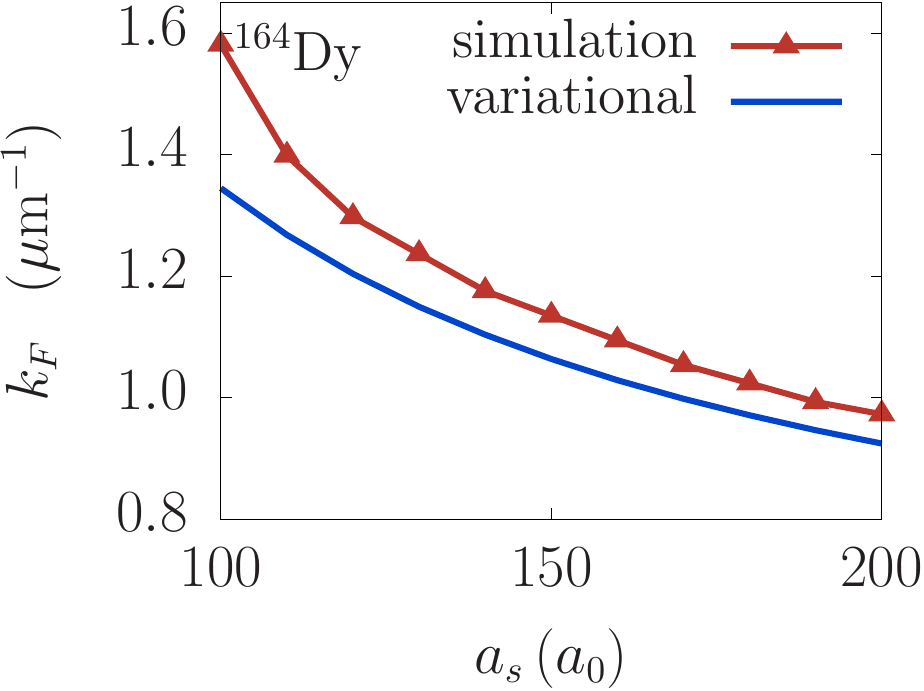}
    \caption{Wave vector of the Faraday waves $k_F$ as a function of the contact interaction strength for a BEC of $^{52}$Cr (\textbf{left}),
     $^{168}$Er (\textbf{middle}),
      and $^{164}$Dy (\textbf{right}),
       for a fixed DDI strength. Red upper triangles were numerically obtained values using the FFT analysis as in Figure~\ref{fig:faraday_fft_k}, and blue lines are the variational results according to Equation (\ref{eq:k_F}).}
    \label{fig:faraday_k_F_acc}
\end{figure}

Now, we turn our attention to spatial features of the Faraday waves. Figure~\ref{fig:faraday_k_F_acc} presents the dependence of the wave vector $k_F$ on the $s$-wave scattering length $a_s$ for all three considered species. We also show the variational results for the dependence $k_F(a_s)$ derived in Section~\ref{sec:variational}. The agreement is very good, with errors of the order of 10--15\%. We stress that the derived variational expression closely follows the numerical results not only by their values, but,   more importantly, also their functional dependence properly.

\begin{figure}[H]
    \centering
        \includegraphics[width=0.31\textwidth]{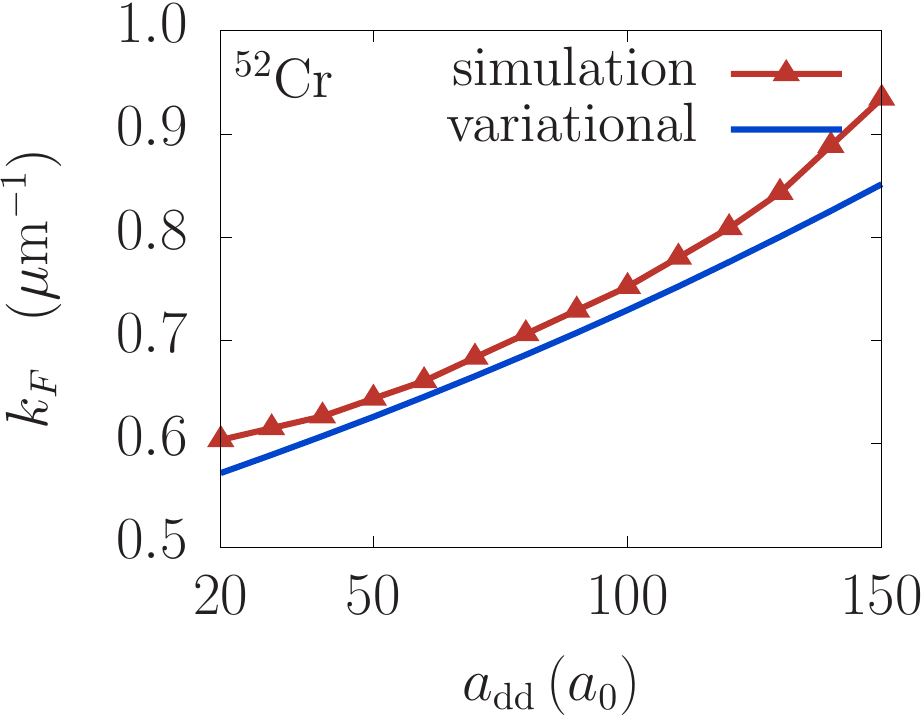}
        \includegraphics[width=0.31\textwidth]{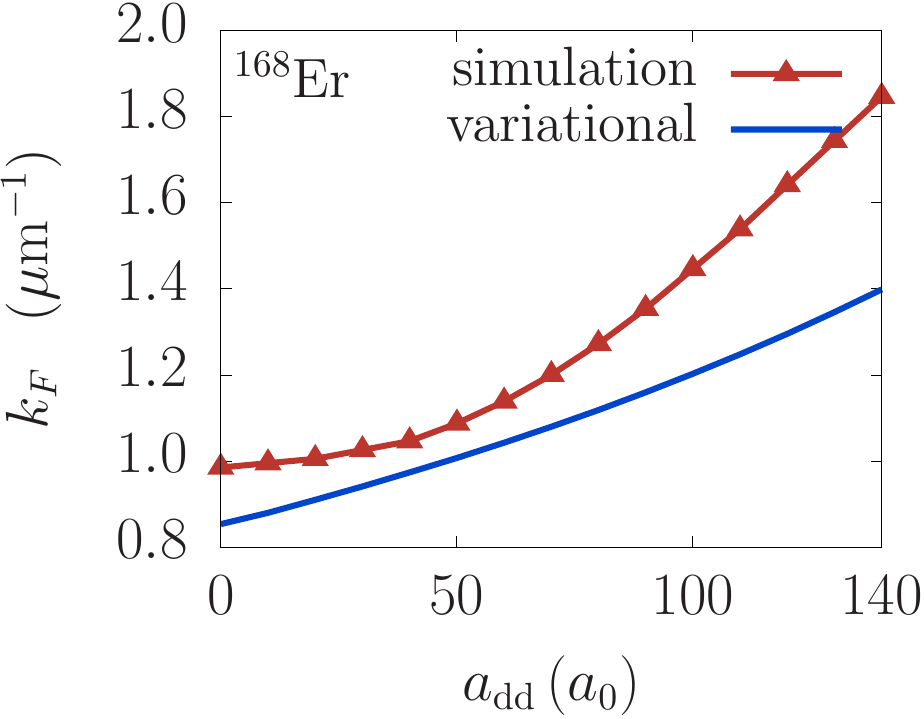}
        \includegraphics[width=0.31\textwidth]{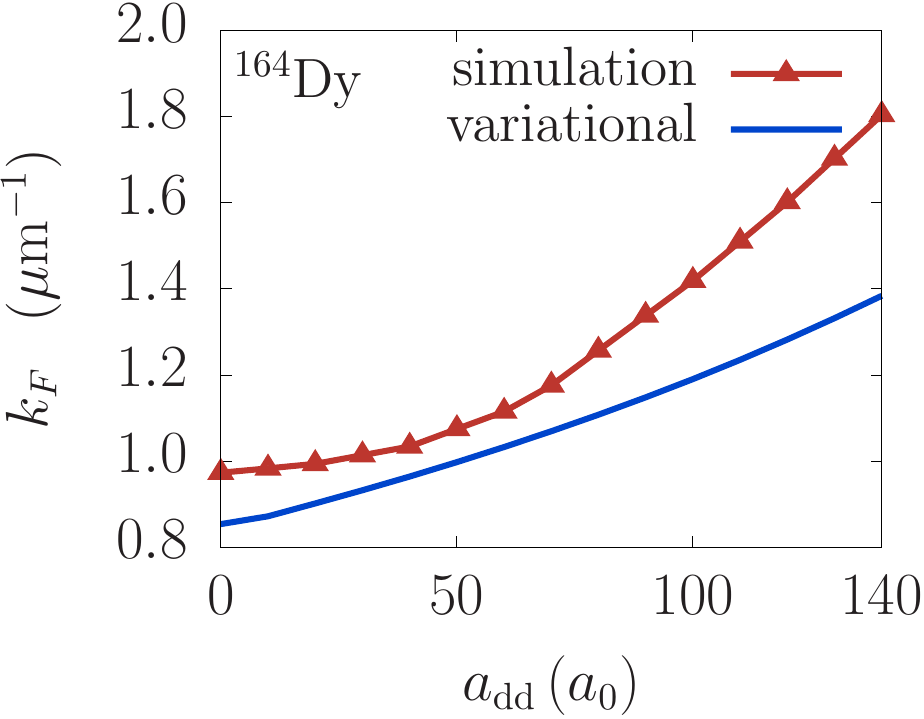}
    \caption{Wave vector of the Faraday waves $k_F$ as a function of the DDI strength for a BEC of $^{52}$Cr (\textbf{left}),
     $^{168}$Er (\textbf{middle}),
      and $^{164}$Dy (\textbf{right}),
       for a fixed contact interaction strength. Red upper triangles represent numerically obtained values using the FFT analysis as in Figure~\ref{fig:faraday_fft_k}, and blue lines are the variational results according to Equation (\ref{eq:k_F}).}
    \label{fig:faraday_k_F_add}
\end{figure}

Next, we studied the effects of the DDI strength for a fixed value of the contact interaction. Figure~\ref{fig:faraday_k_F_add} shows the corresponding dependence of $k_F$ on $a_\mathrm{dd}$. In contrast to the contact interaction dependence, where $k_F$  is a decreasing function of $a_s$, here we see that $k_F$ increases as the DDI strength is increased. Figure~\ref{fig:faraday_k_F_add} also shows the variational results, where the level of agreement with the numerically obtained results is different, with errors as small as 7\% for chromium and up to around 25\% for erbium and dysprosium for largest values of $a_\mathrm{dd}$. Due to complex approximations made in the derivation of variational results, in particular those related to the DDI term, the obtained functional dependence is not as good as in the case of contact interaction, but still provides reasonable estimates of the wave vector values for the Faraday waves.

\section{Resonant Waves}
\label{sec:resonant}

In the presence of interactions, various excitation modes in dipolar BECs are coupled and the energy pumped into the system by periodic driving can be transferred from the driving direction to other, orthogonal directions. In the previous section, we   show this for non-resonant driving, when the harmonic modulation in the radial direction was transferred to the longitudinal direction in the form of Faraday waves, which were the main excitation mode generated. The main distinguishing property of these excitations is halving of the oscillation frequency, i.e., the induced density waves have the frequency $\omega_m / 2$. Next, we studied the other important case, when the modulation frequency is resonant, such that the induced density waves have the same frequency. This happens when $\omega_m$ is close to one of the characteristic frequencies of the system, e.g., one of the frequencies of the collective oscillation modes or one of the trap frequencies. Although Faraday waves and all other collective oscillation modes are also excited in this case, the largest amplitude corresponds to resonant waves with the frequency $\omega_m$. When generated, these resonant waves dominate the behavior of the system and make all other excitations negligible for the dynamics.

\begin{figure}[H]
    \centering
        \includegraphics[width=0.9\textwidth]{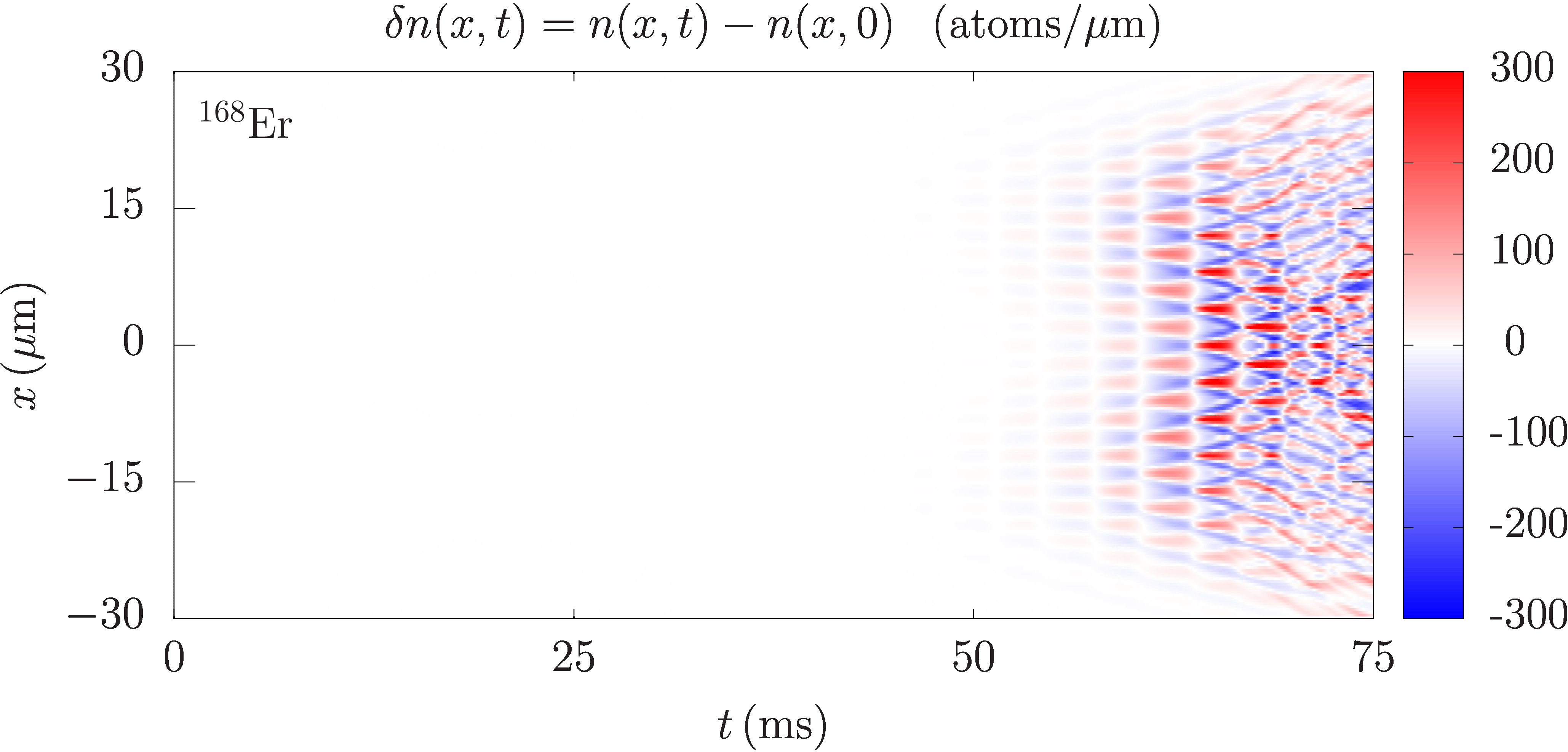}
    \caption{Time evolution of the integrated density profile variation in the weak confinement direction for a BEC of $^{168}$Er, with the modulation frequency  equal to the weak confinement frequency, $\omega_m = \Omega_0$. We observe resonant behavior corresponding to the first harmonic of the resonant frequency $\Omega_0$, which sets in after around $55$~ms.}
    \label{fig:resonant_ndenx_er}
\end{figure}

\begin{figure}[H]
    \centering
        \includegraphics[height=0.3\textwidth]{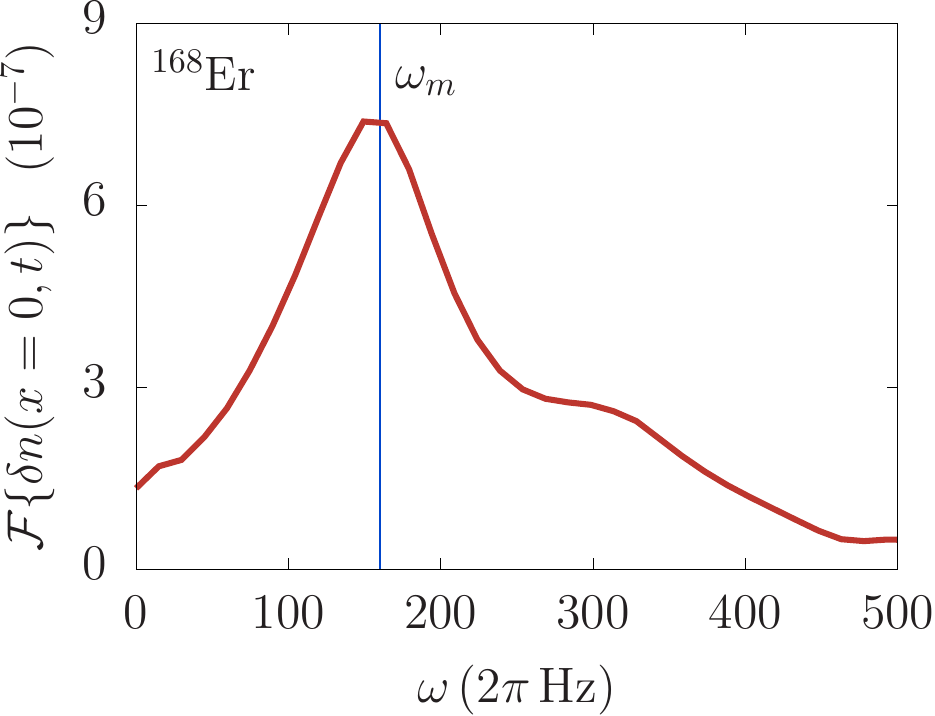}
        \includegraphics[height=0.3\textwidth]{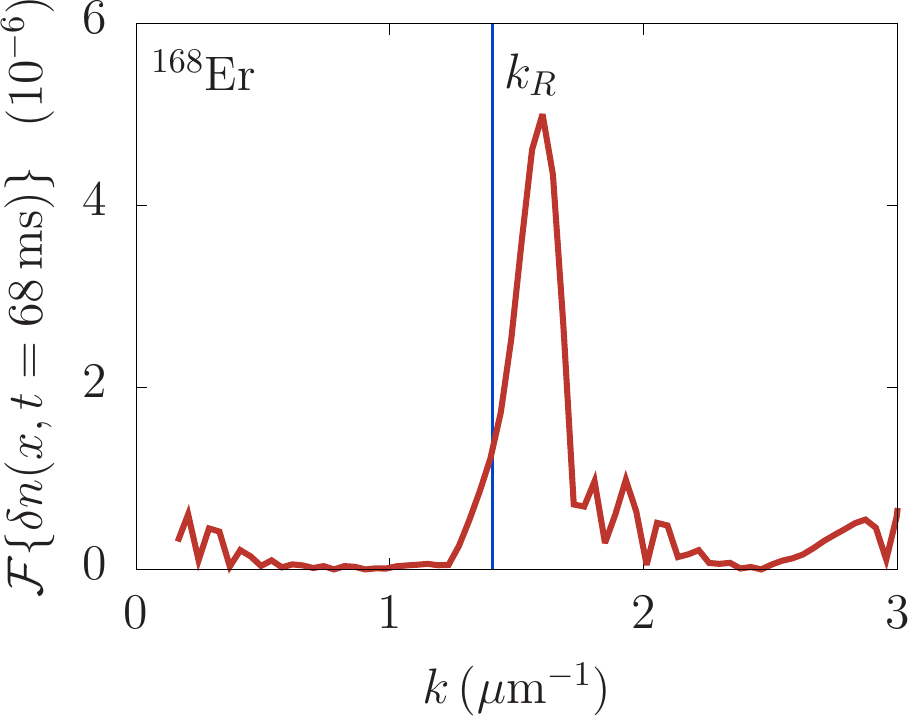}
    \caption{The Fourier spectrum of the integrated 1D density profile variations $\delta n(x, t)$ at the trap center in the time-frequency domain (\textbf{left}),
     and of the density profile variations in $x$ direction $\delta n(x, t = 68~\mathrm{ms})$ in the spatial-frequency domain (\textbf{right})
      of resonant waves for a BEC of $^{168}$Er. Vertical blue line in the left panel represents the modulation frequency $\omega_m$, while in the right panel it corresponds to the theoretical prediction for the wave vector $k_R$ of the resonant waves, Equation (\ref{eq:k_R}).}
    \label{fig:resonant_fft}
\end{figure}

Figure~\ref{fig:resonant_ndenx_er} shows the integrated density profile variation of $^{168}$Er for a resonant wave induced by a harmonic modulation of the radial part of the trapping potential at $\omega_m = \Omega_0$, i.e., when the modulation frequency coincides with the radial trapping frequency. The density waves in this case develop much more quickly than for the non-resonant modulation and are clearly visible already after 55~ms. Due to a violent dynamics that emerges in the system very quickly, it is not easy to estimate the frequency of the waves directly from Figure~\ref{fig:resonant_ndenx_er}, as was possible before. Therefore, we relied on the Fourier analysis in the time-frequency domain, as presented in the left panel of Figure~\ref{fig:resonant_fft}. The obtained FFT spectrum clearly shows that the main excitation mode has the frequency equal to $\omega_m$. We also see that the spectrum is continuous, practically without distinct individual peaks, and only the second harmonic at $2 \omega_m = 321 \times 2 \pi~\mathrm{Hz}$ yields a small local maximum. This demonstrates that the system is far from the regime of small perturbations, where individual excitation modes can be observed.

In the right panel of Figure~\ref{fig:resonant_fft}, we see the Fourier spectrum in the spatial-frequency domain, which yields the wave factor $k_R$ of resonant waves. The FFT results give the value $k_R = 1.59 \, \upmu \mathrm{m}^{-1}$ and the corresponding spatial period $\ell_R = 2 \pi / k_R = 3.95 \, \mathrm{\upmu m}$ for $^{168}$Er. In the figure we also present the variational result $k_R = 1.40 \, \upmu \mathrm{m}^{-1}$, calculated using Equation (\ref{eq:k_R}). The agreement is again quite good, which indicates that the variational approach   developed in this paper can be reliably used not only for the Faraday waves, but also for the resonant waves. 

\begin{figure}[H]
    \centering
        \includegraphics[width=0.4\textwidth]{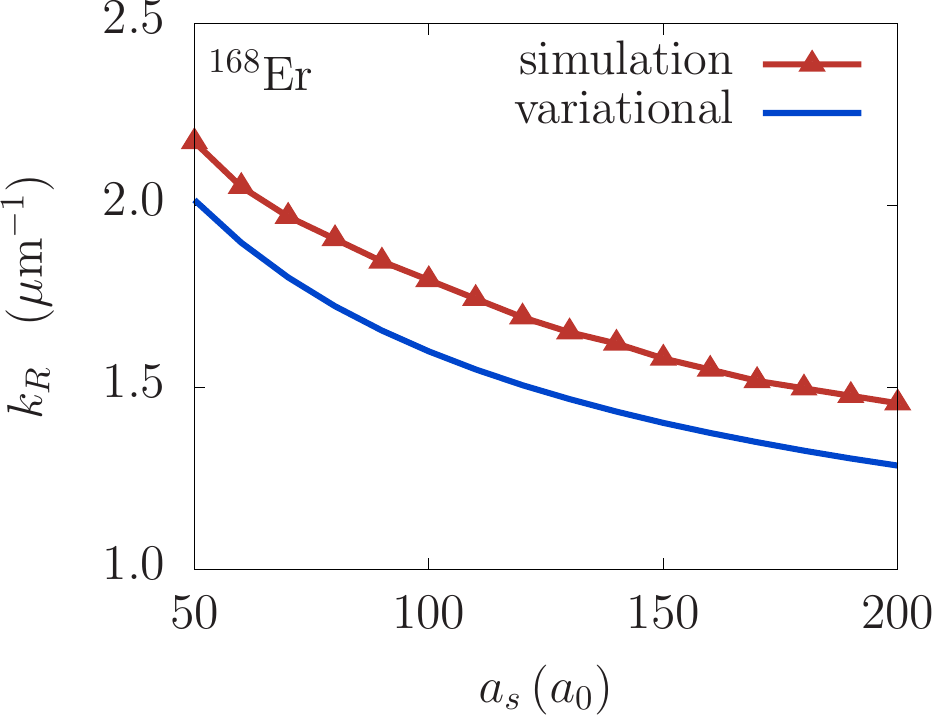}
        \includegraphics[width=0.4\textwidth]{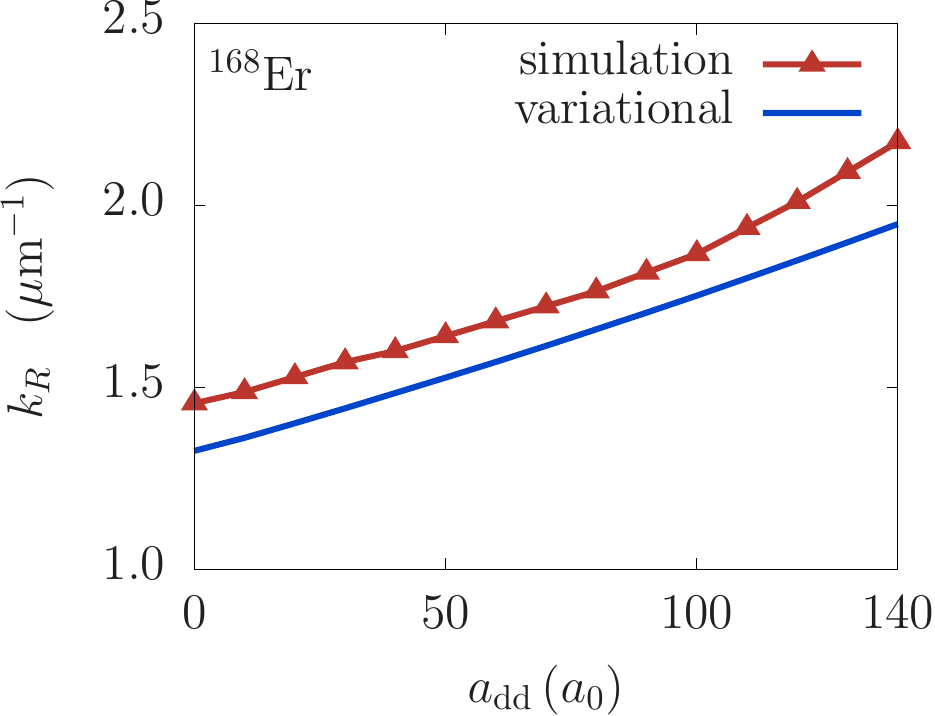}
    \caption{Wave vector of the resonant waves $k_R$ as a function of the contact (\textbf{left})
     and the DDI (\textbf{right})
      strength for a BEC of $^{168}$Er. The results in the left panel are obtained for a fixed DDI strength, and~similarly in the right panel a fixed contact interaction strength is used. In both panels, red upper triangles represent numerically obtained values using the FFT analysis as in the right panel of Figure~\ref{fig:resonant_fft}, and blue lines are the variational results according to Equation (\ref{eq:k_R}).}
    \label{fig:resonant_k_R}
\end{figure}

This can   also be concluded from Figure~\ref{fig:resonant_k_R}, which presents the results for the dependence of the resonant wave vector $k_R$ on the contact and the DDI strength. The agreement between the numerical and variational results is of the order of 10\% over the whole experimentally relevant domain. We see similar behavior for the resonant waves as for the Faraday ones, namely the wave vector decreases as the contact interaction strength increases, while the opposite is true for the DDI. Again,  the functional dependence obtained  from the variational approach properly describes the numerical results, thus confirming that Equation (\ref{eq:k_R}) can be used to calculate spatial period of resonant waves.
\unskip

\begin{figure}[H]
    \centering
        \includegraphics[width=0.8\textwidth]{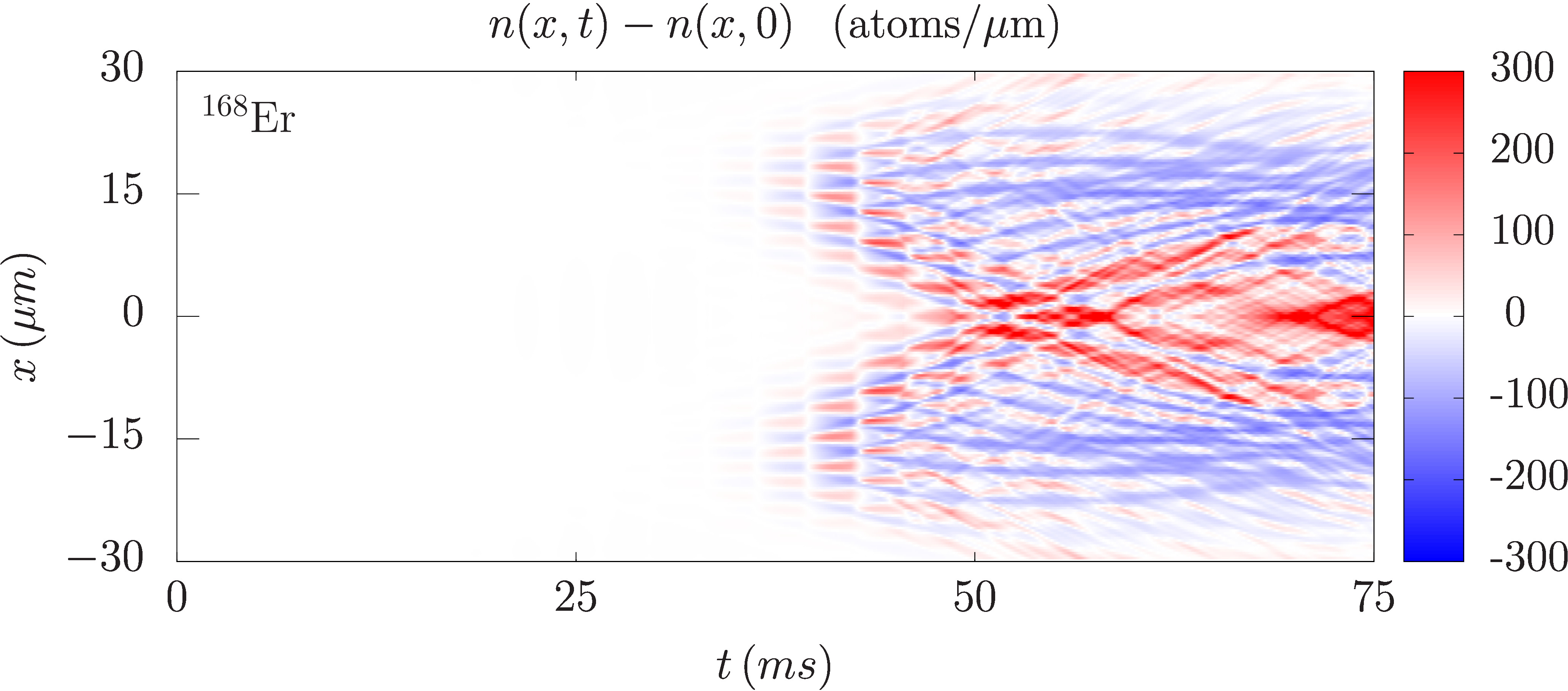}
    \caption{Time evolution of the integrated density profile variation in the weak confinement direction for a BEC of $^{168}$Er. The modulation frequency is equal to twice the weak confinement frequency, $\omega_m = 2\Omega_0$. We observe resonant behavior corresponding to the second harmonic of the resonant frequency $\Omega_0$, which sets in  more quickly  than the first harmonic, already after around $30$~ms.}
    \label{fig:resonant2_ndenx_er}
\end{figure}

It is interesting to note that resonant behavior appears not only under conditions mentioned above, when $\omega_m$ is equal to one of the characteristic frequencies, but also when it matches their higher harmonics. Figure~\ref{fig:resonant2_ndenx_er} illustrates this for $^{168}$Er, which is harmonically modulated at twice the radial trapping frequency, $\omega_m = 2\Omega_0=321 \times 2 \pi$~Hz. In this case, the amplitude of the resonant mode grows even more quickly and significant density variations can be observed already after 30~ms. Therefore, we see that the modulation at the second harmonic yields even more violent dynamics than the first harmonic. The Fourier analysis in the time-frequency domain reveals that the main excitation mode again has a frequency of $\Omega_0$, but the mode at $\omega_m = 2\Omega_0$ is also present. From the experimental point of view, resonant driving is very dangerous and leads to the destruction of the system in a matter of tens of milliseconds. While numerical simulations can be performed for longer time periods, the atoms leave the condensate due to a large, resonant transfer of energy to the system. As the condensate is depleted, the mean-field description of the system breaks down and it can no longer be simulated by the dipolar Gross-Pitaevskii equation.

\section{Conclusions}
\label{sec:conclusions}

We      investigated here the Faraday and resonant density waves in ultracold dipolar Bose-Einstein condensates for experimentally relevant atomic species with the permanent magnetic dipole moment: chromium $^{52}$Cr, erbium $^{168}$Er, and dysprosium $^{164}$Dy. The interplay of the contact and the dipole-dipole interaction in such systems is a hot research topic today, but detailed understanding of their dynamics and even their stability is still lacking. Our results contribute to variational and numerical description of driven dipolar systems and their properties, which are important for ongoing experiments, and will be of particular interest as the strongly dipolar regime becomes experimentally~available.

To describe the dynamics of the Faraday and resonant waves in dipolar BECs, we      relied here on the variational approach introduced in Ref.~\cite{PhysRevA.76.063609} (and references therein), which was already used in various setups~\cite{RomRepPhys.65.820,JPhysB.49.165303,PhysicaA.391.1062,PhysicaA.389.4663,ProcRomAcad.12.209,RomRepPhys.63.1329,ProcRomAcad.14.35,PhysRevA.89.023609,PhysRevE.84.056202,PhysRevA.85.023613}. This approach is based on the Gaussian variational ansatz and includes the condensate widths and the conjugated dynamical phases as parameters. The~ ansatz also includes the density modulations in order to capture the dynamics of density waves. Using our variational approach, the obtained equations for the dynamical evolution of the system are cast into the form of the Mathieu-like differential equation. This allowed us to identify the most unstable solutions of the Mathieu's equation with the Faraday and the resonant waves, which we      observed numerically. Based on this idea, we      derived analytical expressions for the periods of these two types of density waves. Performing the FFT analysis of the results of extensive numerical simulations, we were able to calculate the corresponding periods numerically, as functions of the contact and the dipole-dipole interaction strength. The comparison of variational and numerical results shows very good agreement and demonstrates that the derived analytical expressions provide full understanding of the properties of density waves in dipolar condensates.

In the future, we plan to study onset times for the emergence of Faraday and resonant waves, and in particular the corresponding exponents and their dependence on the contact and the DDI. It~is well known that the periodic driving of a dipolar system may lead to its collapse, and we plan to investigate if recently observed quantum droplets, that appear as a result of stabilization due to quantum fluctuations, may also appear in a scenario which leads to Faraday waves.

\vspace{6pt}
\authorcontributions{Conceptualization, methodology, software, validation, formal analysis, writing---original draft preparation, and writing---review and editing, all authors; investigation, data curation, and visualization, D.V.; and resources, supervision, project administration, and funding acquisition, A.B.}

\funding{This research was funded by the Ministry of Education, Science, and Technological Development of the Republic of Serbia under project ON171017.}

\acknowledgments{We would like to acknowledge inspiring discussions with Vladimir Velji\'{c} and Ivana Vasi\'{c}. Numerical simulations were run on the PARADOX-IV supercomputing facility at the Scientific Computing Laboratory, National Center of Excellence for the Study of Complex Systems, Institute of Physics Belgrade.}

\conflictsofinterest{The authors declare no conflict of interest.} 

\reftitle{References}


\end{document}